\documentclass[aps,twocolumn,PRB,reprint,superscriptaddress]{revtex4-2}

\usepackage{amsfonts}
\usepackage{amsmath}
\usepackage{amssymb}
\usepackage{graphicx}
\usepackage{color}
\usepackage{tensor}
\usepackage{upgreek}
\usepackage{accents}
\usepackage[percent]{overpic}
\usepackage{times}
\usepackage{subfigure}
\usepackage{latexsym}
\usepackage{bm}
\usepackage{anyfontsize}
\usepackage{hyperref}
\hypersetup{colorlinks=true,linkcolor=blue,anchorcolor=blue,citecolor=blue,urlcolor=blue}
\usepackage[svgnames]{xcolor}
\usepackage{mathtools}
\usepackage{multirow}
\usepackage{booktabs}
\usepackage{makecell}
\usepackage{hhline}

\begin{document}
\title{Band-like Exact Zero-energy Andreev Bound States and Superconducting Diode Effect in Mixed \textit{s+p}-wave Josephson Junctions}
\author{Shu-Tong Guan}
\affiliation{National Laboratory of Solid State Microstructures, Department of Physics, Nanjing University, Nanjing 210093, China}
\author{Jin An}
\email{anjin@nju.edu.cn}
\affiliation{National Laboratory of Solid State Microstructures, Department of Physics, Nanjing University, Nanjing 210093, China}
\affiliation{Collaborative Innovation Center of Advanced Microstructures, Nanjing University, Nanjing 210093, China}
\begin{abstract}
Topological Josephson junctions enable nonreciprocal transport involving Majorana fermions (MFs). Here we examine a topological Josephson junction with mixed $s$+$p$-wave pairing, where topological phase transition can be driven by adjusting the ratio between the pairing components. There exist two exact symmetrically positioned zero-energy level crossings for the Andreev-bound states, which can be shifted by external fields, and can be destroyed or recreated in pairs by a time-reversal breaking Zeeman field or inhomogeneities, exhibiting band-like structure. The dependence of the shift on the Zeeman field is linear when the two $p$-wave $\boldsymbol{d}$-vectors on both sides are identical while quadratic when they are distinct. Near the topological phase transition, the topological $p$-wave dominant junctions host MF-induced pronounced superconducting diode effect with high efficiency factor $Q$ up to 30\%, in contrast to the trivial $s$-wave dominant junctions possessing relatively small $Q$. 
\end{abstract}
\date{\today}

\maketitle
\section{introduction }\label{I}
Superconducting diode effect (SDE) \cite{SDE_E7,SDE_E12,SDE_E11,SDE_E3,new20_p,SDE_T26,SDE_T13,SDE_T18,SDE_T10,SDE_79} is a non-reciprocal transport phenomenon in superconductors (SCs), showing direction dependence in critical supercurrents. Besides superconducting chains possessing Cooper pairs with finite momenta  \cite{SDE_Fu_liang,SDE_T27,SDE_T21,SDE_T7,SDE_T9,SDE_E5,ABS_712}, SDE can manifest itself in Josephson junctions \cite{SDE_T9,SDE_E5,ABS_712,SDE_E13,SDE_E4,SDE_E10,SDE_75,SDE_E72,J_78,SDE_E14,SDE_E2,SDE_E15,SDE_T25,SDE_T20,SDE_T12,SDE_T23,SDE_T19,SDE_T3,16_ref3,17_1_ref3,SDE_T5,17_ref3}, which generally break both time-reversal ($\boldsymbol{\mathcal{T}}$) and spatial inversion ($\boldsymbol{\mathcal{P}}$) symmetries. For short junctions, the supercurrent is mainly carried by the Andreev-bound states (ABSs). Particularly in the corresponding topological Josephson junctions there exist MF-induced zero-energy ABSs (ZEABSs) at $\pi$ phase difference \cite{Fu_Liang,new_12_Fu_Liang,JJ_MZM_E0_5,S+P,JJ_MZM_E0_2,new1_nan0_J_ABS0,JJ_MZM_E0_9,Nano_pi_1,MZM4_J,Nano_pi_2,Nano_pi_3,new_11_ABS_MZM,GREEN4,JJ_MZM_E0_7,JJ_MZM_E0_1}, at which current abruptly changes and critical currents are formed in the vicinity. The connections between the SDE together with its efficiency factor and the interface MFs are still under debate \cite{SDE_T14,SDE_T16,SDE_T17,SDE_T1,SDE_T2,SDE_Nagaosa}. In most of topological Josephson junctions including both theoretical proposals and experiments \cite{Fu_Liang,MZM4_J,Nano_pi_2,Nano_pi_3,SDE_T2,SDE_Nagaosa,JJ_MZM_E0_6,MZM12_J,MZM13_J,MZM7_J,MZM8_pi}, nanowires with spin-orbit coupling (SOC) and superconducting proximity are investigated, where an effective $p$-wave is formed in the nanowires. However, as one of the characteristic feature of a $p$-wave SC, the $\boldsymbol{d}$-vector is relatively locked in the effective $p$-wave nanowires. How this degree of freedom affect the topological Josephson junctions including their MF-induced ZEABSs \cite{d_vector_8} and SDE \cite{SDE_T8} has rarely been studied. Since the orientation of $\boldsymbol{d}$ is intrinsically relevant to spin-dependent pairing phase \cite{d_vector3,d_vector_8,d_vector2,GREEN4}, its manipulation is expected to be capable of capturing the characteristic feature of the induced interface MFs and relevant transport properties.

In this paper we examine a mixed $s$+$p$-wave Josephson junction, where SCs on both sides of the normal link may take distinct $p$-wave $\boldsymbol{d}$-vectors, and topological phase transition can be driven by adjusting the ratio between different pairing components. We found that while a Zeeman field $\boldsymbol{h}$ can always induce a pair of symmetrically positioned MF relevant ZEABSs, their position dependence on $h$ is linear as the two $\boldsymbol{d}$-vectors are identical while is quadratic as they are distinct. The ZEABSs can be destroyed or recreated in pairs but only at $\varphi=0$ or $\pi$, with $\varphi$ the phase difference, and their existence region can thus show a band-like structure. By symmetry analysis we found when the two $\boldsymbol{d}$-vectors are anti-parallel or when they are identical and $\boldsymbol{d\cdot h}=0$, SDE is forbidden. Otherwise, both topological $p$-wave dominant phase and trivial $s$-wave dominant one can host SDE, but near the topological phase transition, they can be distinguished from each other by their efficiency factors. This phenomena may thus provide a potential application of Josephson diodes as topological phase transition indicators. 

This paper is organized as follows. In Sec.~\ref{II}, we analyze in $p$-wave dominant junctions the exact ZEABSs, demostrating their band-like structure under external fields. In Sec.~\ref{III}, we discuss the existence of SDE through symmetry analysis, no matter whether the system is topologically nontrivial or not. In Sec.~\ref{IV}, we discuss the current-phase relations indicating SDE, distinguishing the topological nontivial junctions from trivial ones by their efficiency factors. In Sec.~\ref{V}, We extend our results to the junctions with distinct $\boldsymbol{d}$-vectors.

\begin{figure}[!t]
  \begin{center}
	\includegraphics[width=8.6cm,height=3.3115cm]{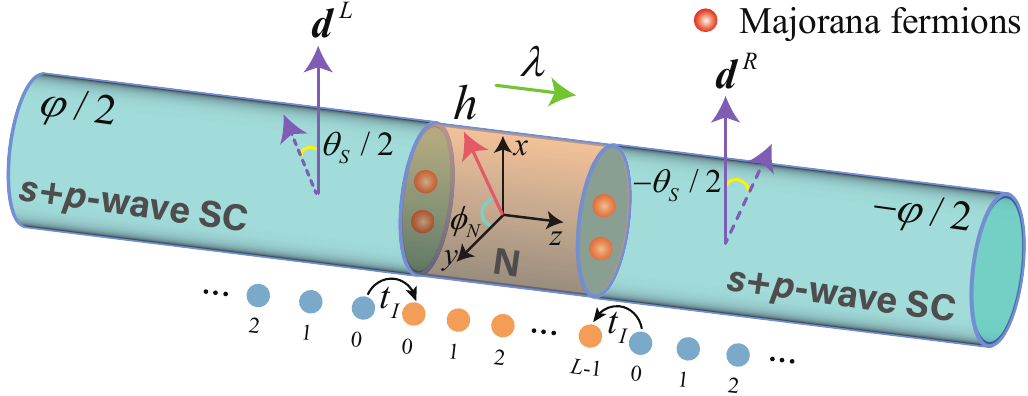}
  \end{center}
  \vspace{-0.2cm}
  \caption{Josephson Junction composed of the mixed $s$+$p$-wave pairing topological SCs, where the phase difference is $\varphi$, and the $p$-wave $\boldsymbol{d}$-vectors in both SCs, being within the $xy$ plane, can be identical or be differing by $\theta_S$ angle. A Kramers pair of MFs occurs at each interface. In the normal link with length $L$, a Zeeman field $\boldsymbol{h}$ and Rashba SOC $\lambda$ are taken into account.} 
\label{fig1} 
\end{figure}

\section{Exact ZEABSs and their band-like structure}\label{II}
We study a 1D lattice model as schematically illustrated in Fig.~\ref{fig1}. Although a pairing with mixed parity is allowed only if $\boldsymbol{\mathcal{P}}$ symmetry is broken, to extract the essential physics in the junction, we neglect a possible small SOC in the SC for simplicity. Our model Hamiltonians can be given by:
	\begin{align}
    H_{S}
    &=\sum_{k}c^{\dagger}_{k}\xi_k c_{k}+\sum_{k}\left(\Delta_s  c^{\dagger}_{k \uparrow}c^{\dagger}_{-k \downarrow}+\text{h.c}\right) \nonumber\\ 
    &+\frac{1}{2}\sum_{k}\left[\Delta_p \sin{k}\left(c^{\dagger}_{k \uparrow}c^{\dagger}_{-k \uparrow}-c^{\dagger}_{k \downarrow}c^{\dagger}_{-k \downarrow}\right)+\text{h.c}\right],\label{eq1}\\ 
 H_N&=\sum_{k}c^{\dagger}_{k}\left(\xi_k-\boldsymbol{h} \cdot \boldsymbol{\sigma}+\lambda\sin{k}\sigma_z\right) c_{k}, \label{eq2}
 \end{align}
where $ c_k=(c_{k \uparrow},\,c_{k \downarrow})^{T}$ with $\boldsymbol{\sigma}$ the spin Pauli matrices, $\Delta_s$ ($\Delta_p$) is the on-site $s$-wave (the nearest-neighbor $p$-wave) pairing potential, and $\xi_k=-2\cos{k}-\mu$ with $\mu$ the chemical potential. In the normal link, a generic Zeeman field $\boldsymbol{h}$ and a Rashba SOC $\lambda$ are taken into account. In the weak-pairing limit, whether a 1D SC is topological depends on the sign of the product of the order parameter at two Fermi points: $\Delta_{k_F}\Delta_{-k_F}$, where $\Delta_{k}=\Delta_s+\Delta_p \sin{k}$. So the SC with mixed parity is topologically nontrivial when the $p$-wave pairing is dominant, namely, when $\Delta_s$$<$$\Bar{\Delta}_p\equiv\Delta_p \sin{k_F}$, otherwise it's trivial. The $p$-wave $\boldsymbol d$-vector $\boldsymbol{d_k}$, which appears in the pairing matrix as $(\boldsymbol{d_k} \cdot \boldsymbol{\sigma})i\sigma_y$, is chosen here to be along $\boldsymbol{x}$ for convenience: $\boldsymbol{d_k}=-\Delta_p\sin k(1,0,0)$.

\begin{figure}[t]
  \begin{center}
	\includegraphics[width=8.5cm,height=6.578cm]{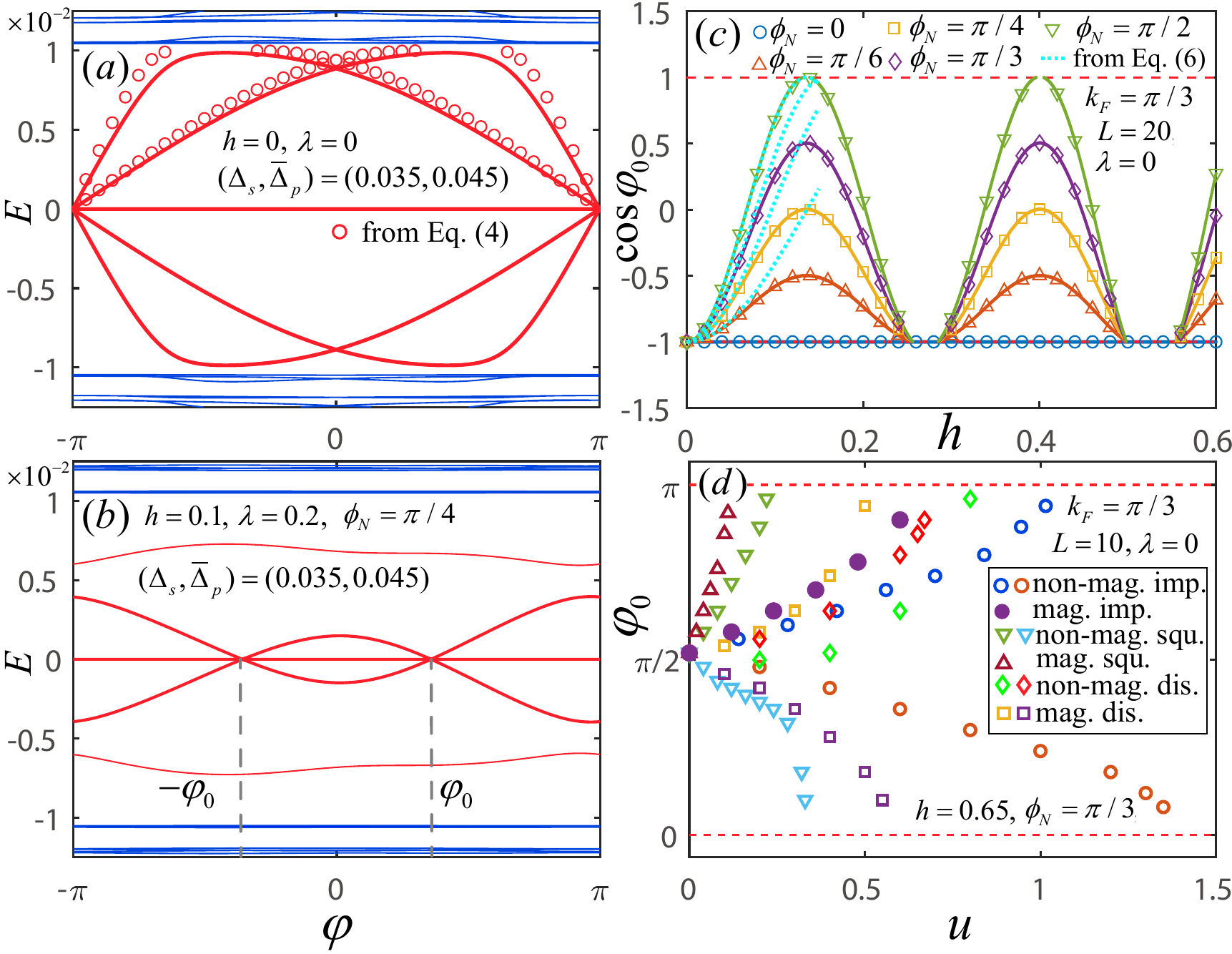}
  \end{center}
  \vspace{-0.5cm}
  \caption{ZEABSs and their band-like structure for a $p$-wave dominant mixed pairing Josephson junction. (a)-(b): ABS spectra as functions of $\varphi$, where the zero-energy flat bands correspond to the MFs at open ends of the two SCs, and $\pm\varphi_0$ mark out the exact ZEABSs induced by the MFs at two interfaces. Other parameters: $L=10$, $k_F=\pi/6$, $t_I=-0.6$. (c) Band structure of the ZEABSs as functions of the magnitude $h$ for different orientations of the Zeeman field. (d) Shift of $\varphi_0$ with the increase of inhomogeneity amplitude, including the on-site nonmagnetic/magnetic impurity potentials (open/solid circles) $\pm u$$/u$ at site $j=6$, square potentials (down-triangles) $\pm u$ within $2\leq j\leq 7$ or their magnetic counterpart (up-triangles), and four randomized potentials within $[-u,u]$ (diamonds and squares).} 
\label{fig2}
\end{figure}

When $L\ll\xi$ with $\xi$ the superconducting coherence length, the Josephson current is mainly contributed by the ABSs, which are formed within the normal region, due to multiple particle or hole reflections between the two NS interfaces. For a homogeneous link, one can invoke the formula below to find them \cite{Beenakker}:  
\begin{equation}
    \begin{aligned}
    \text{det}\left(\mathbf{1}-T_L \boldsymbol{r}_1 T_L \boldsymbol{r}_2\right)=0,
    \end{aligned}
    \label{eq3}
\end{equation}
where $T_L$ is the propagation matrix between the two interfaces and $\boldsymbol{r}_1$ ($\boldsymbol{r}_2$) is the unitary refection amplitude matrix incident from the normal region for the left (right) interface. See Appendix~\ref{C} for detail. By independently matching the boundary conditions (BCs) for the corresponding scattering state at each interface to find $\boldsymbol{r}_i$, we obtain the ABSs, which can be determined by (see Appendix~\ref{C}): 
\begin{equation}
\begin{aligned}
\cos(\theta_+ + \theta_-)-\cos(\pm\varphi+\theta_+ - \theta_-) \qquad \qquad \qquad \qquad \quad\\
=4\sin^2\frac{(\theta_+ + \theta_-)}{2} \{\frac{(1-t_I^2)^2}{(\hbar v_F t_I)^2}+\frac{(1-t_I^2)^4}{(\hbar  v_F t_I)^4 }+\frac{(1-t_I^2)^2}{(\hbar  v_F t_I)^4 }\times \\
[-\cos 2k_F(L+1)+ 2t_I^2\cos 2k_F L -t_I^4 \cos 2k_F(L-1) ]\},
\end{aligned}
\label{eq4}
\end{equation}
where  $e^{i\theta_\pm}$$=$$\left(E-i({\Delta_{\pm k_F}^{2}-E^2})^{1/2}\right)/{\Delta_{\pm k_F}}$, $t_I$ is the interface hopping integral, and $\hbar v_F=2\sin k_F$. This equation gives the ABS spectrum within $E$$<$$\Delta_\text{min}$$\equiv$$\mid$$\Delta_{-k_F}$$\mid$ in the absence of the Zeeman field and SOC. A particular situation is $t_I^2=1$, which causes $\cos(\theta_+ + \theta_-)-\cos(\pm\varphi+\theta_+ - \theta_-)=0$. It obscures the difference between the topologically trivial and nontrivial SCs, especially when $E=0$, which is always a solution at $\varphi=\pi$, since $\theta_-=\theta_+=-\pi/2$ for the $s$-wave dominant case while $\theta_-=-\theta_+=\pi/2$ for the $p$-wave dominant one. This uncertainty can be distinguished by simply considering the interface roughness: the deviation of $t_I^2$ from $1$. With $t_I^2\neq1$, the $s$-wave dominant topologically trivial SC can have a ZEABS at $\varphi=\pi$ only when the square bracket in Eq.~(\ref{eq4}) becomes accidentally zero. Even in this situation, this state can still easily becomes avoided crossing under parameter change. However, when the SC is $p$-wave dominant, these always exists a Kramers pair of MFs localized at each interface, inducing doubly degenerate exact level crossing at $\varphi=\pi$, as shown in Fig.~\ref{fig2}(a). Although Eq.(\ref{eq4}) is approximate, this level crossing is exact and protected due to the local fermion parity conservation \cite{Fu_Liang,new_12_Fu_Liang,fermion-parity} of our model Hamiltonian. 
 
In the presence of a Zeeman field and Rashba SOC, as long as their strengths are small enough, the exact crossing will not be destroyed but would split and be shifted symmetrically to new locations at $\varphi=\pm\varphi_0$. In contrast to the well studied nanowire Josephson junctions, where the effective $p$-wave pairing induces a single level crossing at $\varphi=\pi$ \cite{Nano_pi_1,Nano_pi_2,Nano_pi_3} which cannot be removed by fermion-parity conserving external fields or interactions, the two new crossings here can be removed in pairs by a $\boldsymbol{\mathcal{T }}$-breaking perturbation. We take two exact analytical results as illustration: one is a purely SOC case without Zeeman field, where the crossings are always being protected, with $\varphi_0$ fixed at $\pi$ as the SOC is parallel to $\boldsymbol{d}_k$, while $\varphi_0=\pi+\lambda (L-1)$ (see Appendix~\ref{D}) as it is perpendicular to $\boldsymbol{d}_k$. The other is a purely Zeeman-field case without SOC, where $\boldsymbol{h}$ is parallel to $\boldsymbol{d}_k$ ($\phi_N$$=$$\pi/2$, see Fig.~\ref{fig1}), and the new location $\varphi_0$ obeys (see Appendix~\ref{D}): 
\begin{equation}
    \begin{aligned}
    \cos{ \varphi_0}=-\cos{\Delta k_F L}+\tan^2{ \Theta }(\cos{2 k_F L}-\cos{\Delta k_F L}),
    \end{aligned}
    \label{eq5}
\end{equation}
where $\Delta k_F=k_F^+-k_F^-$ with $k_F^{\pm}$ the two spin species' Fermi wave vectors in the normal region, and $\tan{ \Theta }=(\eta^{1/2}-\eta^{-1/2})/2$ with $\eta=\tan{(k_{F}^+/2)}/\tan{(k_{F}^-/2)}$. When Eq.~(\ref{eq5}) has no solution, it means the absence of the exact ZEABSs, but Eq.~(\ref{eq5}) does always admit a solution when $h\ll1$, giving rise to the following linear dependence of $\varphi_0$ on $h$ extended for a generic $\phi_N$ (see Appendix~\ref{D}): 
\begin{equation}
\begin{aligned}
\varphi_0&=\pi+\frac{h \sin{\phi_N}}{\sin^2{k_F}}\sqrt{( L\sin{k_F})^2-\sin^2{(k_FL)}}.
\end{aligned}
\label{eq6}
\end{equation}
For some relative large $h$ and $\lambda$, exact ZEABSs can still exist, as shown in Fig.~\ref{fig2}(b), where two symmetrically positioned level crossings are found at $\varphi=\pm\varphi_0$, although the total energy spectrum is not symmetric at all. Actually, the righthand side of Eq.~(\ref{eq5}) exhibits an oscillating behavior as a function of $h$ with some of their extremum absolute values larger than $1$. The existence region of the exact ZEABSs and the no-solution region occur alternatively, giving rise to a band-like structure for the former, as shown in Fig.~\ref{fig2}(c). Band edges are always located at $\varphi_0=\pi$ for most of $\phi_N$ while at $\varphi_0=0$ or $\pi$ for $\phi_N\approx\pi/2$, at which the ZEABSs are removed in pairs, keeping fermion parity conserved.  

The locations $\pm\varphi_0$ of the exact crossings are independent of the details of the SC such as $\Delta_s$, $\Delta_p$, and that of the interfaces such as $t_I$. Below we shall demonstrate rigorously that according to Eq.~(\ref{eq3}), this independence actually relies on the independence of $\boldsymbol{r}_i(0)$ at $E=0$ on the details. To find $\boldsymbol{r}_i(0)$ exactly, one has to treat the BCs at each interface very carefully. For example, the BCs at the left interface are:
\begin{equation}
    \begin{aligned} 
    \begin{dcases}
     (\tau_z-i\lambda\sigma_z/2) \Psi_N(-1)&=-t_I \tau_z \Psi_S(0),\\
    (\tau_z-i\Delta_p\sigma_x\tilde \tau_1/2) \Psi_S(-1)&=-t_I \tau_z \Psi_N(0).
    \end{dcases}
    \end{aligned}
    \label{eq7}
\end{equation}
where $\tilde \tau_1=\cos{(\varphi/2)}\tau_x-\sin{(\varphi/2)}\tau_y$ with $\boldsymbol{\tau}$ the particle-hole Pauli matrices. See Fig.~\ref{fig1}. Here $\boldsymbol{d}_k$ is chosen along $\bm{z}$ for convenience. These BCs are actually the Schrodinger equations for the nearby sites at the left interface, being simplified by introducing auxiliary sites $j$$=$$-1$ for both regions \cite{Zhou_Yao,Liu_Chun_Chi}. $\Psi_N(j)$ is the incident and reflected wave function containing $\boldsymbol{r}_1(0)$ entries, while $\Psi_S(j)$ is the transmitted wave function given by $\Psi_S(j)=\bm{u}_+ t_{e} e^{iq_+ j}+\bm{u}_- t_{h} e^{-iq_- j}+\bm{u}'_+ t_{e}' e^{iq'_+ j}+\bm{u}'_- t_{h}' e^{-iq'_- j}$, where $t_{e/h}$ and $t'_{e/h}$ are electron/hole-like transmission amplitudes, with $\pm q_\pm$ and $\pm q'_\pm$ the corresponding complex wave vectors. Accordingly $\bm{u}_\pm$ and $\bm{u}'_\pm$ are the eigenmodes: $\bm{u}_\pm=(1,0,0,\pm \exp i(\theta_\mp-\varphi/2))^T$, and $\bm{u}'_\pm=(0,1,\pm \exp i(\theta_\mp-\varphi/2),0)^T$. At $E=0$, a $p$-wave dominant SC means $\theta_-=-\theta_+=\pi/2$, indicating the coalesce of the eigenmodes: $\bm{u}_+=\bm{u}_-\equiv\bm{u}$ and $\bm{u}'_+=\bm{u}'_-\equiv\bm{u}'$. $\Psi_S(j)$ then becomes the superposition of the two localized MFs at the interface. Since both the coalesced eigenvectors $\bm{u}$ and $\bm{u}'$ are also the eigenvectors of $i\sigma_x\tilde\tau_z \tau_1$ with the same eigenvalue $-1$, the wave functions satisfy $(1-i\lambda\sigma_z\tau_z/2)\Psi_N(-1)=-t_I\Psi_S(0)$ and $\Psi_N(0)=-t_I^{-1}(1+\Delta_p/2)\Psi_S^1(-1)$. Therefore, only the four combined parameters $t_I(t_e+t_h)$, $t_I(t_{e}'+t_{h}')$, $t_I^{-1}(1+\Delta_p/2)(t_ee^{-iq_+}+t_h e^{iq_-})$ and $t_I^{-1}(1+\Delta_p/2)(t_{e}' e^{-iq'_+}+t_{h}' e^{iq'_-})$ are entered as a whole into the boundary conditions, eliminating the effect of the detailed values of $\Delta_s$, $\Delta_p$ and $t_I$ on the reflection amplitude $\boldsymbol{r}_1(0)$. Due to the same reason, $\boldsymbol{r}_2(0)$ is independent of these detailed parameters and so do the ZEABSs if they exist. Since this feature is absent for the nanowire Josephson junctions, the above discussion can be viewed as another unique proof of the exactness of the level crossings for our $p$-wave dominant junction.

\begin{table}[t]
\begin{center}
\caption{Cases host SDE in a $s$+$p$-wave Josephson junction, where the Rashba SOC in the normal link is along $\bm{z}$ direction.}
\label{tab1}
\begin{tabular}{ >{\centering\arraybackslash}m{0.38\linewidth} |>{\centering\arraybackslash}m{0.59\linewidth} }
            \hhline{==}
            purely $p$-wave with SOC  & mixed $s$+$p$-wave \\ \hhline{=|=} 
            $h_z\ne0 $ & $\lambda\ne 0$:\;\;$h_z\ne0$;\;\;or $d_z \ne0$ and $\boldsymbol{d}_k\nparallel  \boldsymbol{z}$ \\  \hline 
            or $d_z\ne0$ and $\boldsymbol{d}_k\cdot\boldsymbol{h}\ne 0$ & or $\boldsymbol{d}_k\cdot\boldsymbol{h}\ne 0$ \\ \hhline{==}
        \end{tabular}
\end{center}
\vspace{-0.2cm}
\end{table}

We now examine for the $p$-wave dominant case the robustness of the ZEABSs against the inhomogeneity in the normal region. Here the normal link can be viewed as a scattering region described by a unitary scattering matrix $S_N(\{u_\alpha\})$, with $\{u_\alpha\}$ a series of parameters including also possible spatial fluctuations of the Zeeman field and SOC, describing all kinds of the inhomogeneities. $S_N(\{u_\alpha\})$ must be a smooth function of the parameters. The formula Eq.~(\ref{eq3}) determining the ABSs becomes: $\text{det}(\mathbf{1}-S_N\boldsymbol{r})=0$, with $\boldsymbol{r}=$diag$(\boldsymbol{r}_1,\boldsymbol{r}_2)$. Without solving this equation, it must be an implicit real function of $\cos\varphi$ and $\{u_\alpha\}$. The dependence not on $\varphi$ but on $\cos\varphi$ is because physically $\varphi$ and $\varphi+2\pi$ are gauge equivalent. Assuming $\{u_\alpha\}=0$ initially, we start from a Josephson junction system with a pair of exact ZEABSs located at $\varphi=\pm\varphi_0$, where $\cos\varphi_0\neq\pm1$. These two ZEABSs must be robust when the inhomogeneity parameters $\{u_\alpha\}$ are slowly turned on, as long as they are sufficiently small, since the implicit function is analytic, introduction of a small inhomogeneity cannot remove the $E=0$ solutions at $\varphi=\pm\varphi_0$, but shift their positions instead. With the increase of inhomogeneity amplitude, the exact $E=0$ solutions would be removed only at $\cos\varphi_0=\pm1$. In Fig.~\ref{fig2}(d), we demonstrate the robustness of a junction state with its initial $\varphi_0\approx\pi/2$ (with a homogeneous $\boldsymbol{\mathcal{T }}$-breaking Zeeman field) against different inhomogeneities, with the ZEABSs ultimately removed in pairs at $\varphi_0=0$ or $\pi$ \cite{SM1}, while starting from a junction state with initial $\varphi_0=\pi$ free of Zeeman field, the ZEABSs can only be removed by $\boldsymbol{\mathcal{T }}$-breaking inhomogeneities or perturbations.

\section{SDE related symmetry analysis}\label{III}
The SDE of a Josephson junction relies on whether its current-phase relation (CPR) $I( \varphi)$ is non-centrosymmetric. Generally the emergence of SDE must break both $\boldsymbol{\mathcal{T }}$ and $\boldsymbol{\mathcal{P}}$ symmetries. But violation of both symmetries is insufficient. To concentrate to the mixed $s$+$p$-wave junctions hosting SDE, we employ symmetry analysis to exclude the ordinary cases with: $I( \varphi)=-I(-\varphi)$. Here we shall discuss a more general situation where $\boldsymbol{h}$ and $\boldsymbol{d}$ are not constrained to be within the $xy$ plane. Individual $\boldsymbol{\mathcal{T }}$ or $\boldsymbol{\mathcal{P}}$ operation (with respect to the center $\mathcal{C}$ of the normal link) for the whole junction system yields the following relation:
    \begin{align}
        I(-\varphi,\boldsymbol d,-\boldsymbol{h},\lambda)=-I_0,\;\boldsymbol{\mathcal{T }},\label{eq8}\\
        I(-\varphi,-\boldsymbol d,\boldsymbol{h},-\lambda)=-I_0,\;\boldsymbol{\mathcal{P}},\label{eq9}
    \end{align}  
\begin{figure}[!bt]
  \begin{center}
	\includegraphics[width=8.5cm,height=   9.52215 cm]{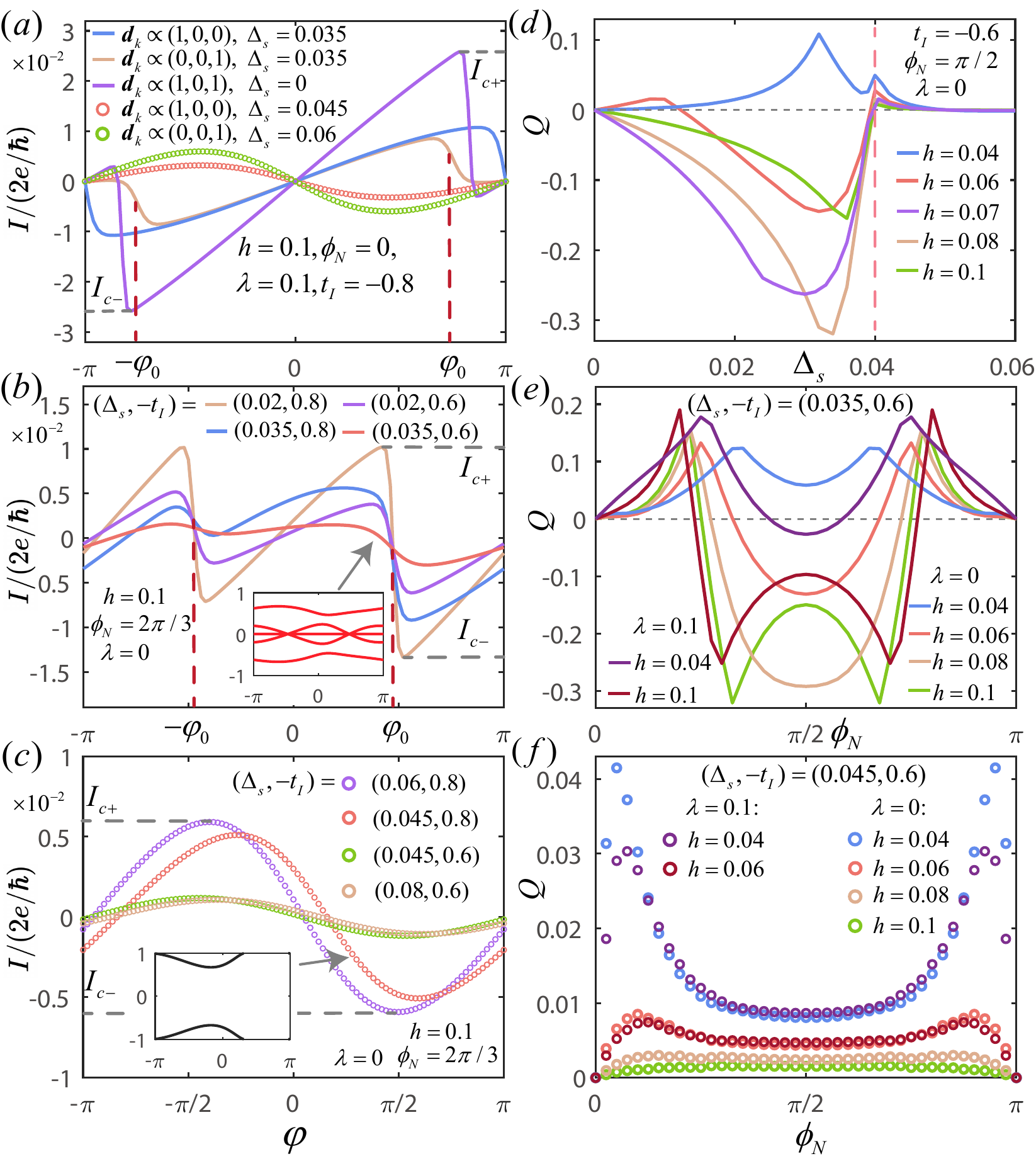}
  \end{center}
  \vspace{-0.6cm}
  \caption{CPRs and efficiency factors $Q$, where parameters are chosen to fix $\Delta_s+\Bar{\Delta}_p$ to be 0.08, with $\Delta_s<(>)0.04$ denoting the p~(s)-wave dominant cases. CPRs of typical cases without SDE for (a) two $p$-wave dominant (solid lines) and two $s$-wave dominant (open circles) cases, together with a purely $p$-wave case. CPRs with SDE for (b) the topologically nontrivial phases and (c) their trivial counterparts. The insets give the ABS spectra for the corresponding data. (d) $Q$ as a function of $\Delta_s$ for different $h$, where the dashed line indicates the topological phase transition at $0.04$. $Q$ versus the orientation $\phi_N$ of the Zeeman field with different $h$ and SOC $\lambda$ in (e) a $p$-wave dominant phase, and in (f) a $s$-wave dominant phase. Here the $\boldsymbol{d}$-vector is along $\bm{x}$ except (a) and common parameters: $k_F=\pi/6$, $L=10$, $k_B T=4\times 10^{-4}$.} 
   \label{fig3}
\end{figure}
where $I_0\equiv I( \varphi,\boldsymbol d,\boldsymbol{h},\lambda)$. Because a mixed $s$+$p$-wave SC has already broken $\boldsymbol{\mathcal{P}}$, SOC is not a necessary ingredient governing the SDE relevant phenomena. Another two independent relations can be obtained by performing $\pi$ rotations around $\mathcal{C}$: \begin{align}
    &I(-\varphi,-d_x,d_y,d_z,h_x,-h_y,-h_z,\lambda )=-I_0,\;\boldsymbol{\mathcal{R }} _x(\pi), \label{eq10}\\
    &I(-\varphi,d_x,-d_y,d_z,-h_x,h_y,-h_z,\lambda )=-I_0,\;\boldsymbol{\mathcal{R }} _y(\pi). \label{eq11}
\end{align}
Combination of Eq.~(\ref{eq9})-(\ref{eq11}) leads to three simplified equations: $d_x h_x$$:$$+$, $d_y h_y$$:$$+$, $d_z h_z \lambda$$:$$+$, with Eq.~(\ref{eq8}) restated as $h_x h_y h_z$$:$$-$, where $x_1x_2x_3$$:$$\pm$ means $I(\varphi,-x_1,-x_2,-x_3)=\pm I(\pm\varphi,x_1,x_2,x_3)$ and $x_1x_2$$:$$\pm$ has a similar meaning. When $\bm{h}$$\perp$$\boldsymbol{d}_k$ we found if $\lambda=0$, or $h_z=0$ with $\boldsymbol{d}_k$ along or perpendicular to $\bm{z}$, SDE is forbidden. For the purely $p$-wave case with SOC, since a minus sign of $\boldsymbol{d}_k$ can be absorbed by the superconducting phase, we have an additional constraint: $d_x d_y d_z$$:$$+$, which means that when $h_z=0$, if $\bm{h}$$\perp$$\boldsymbol{d}_k$, or $d_z=0$, SDE is forbidden. While for the trivial purely $s$-wave case with SOC, $\boldsymbol{d}_k=0$, if $h_z=0$, SDE is forbidden. To seek the situations with SDE, we need only to avoid the above forbidden ones and in most of the remaining cases they have been confirmed to support SDE, which are summarized in Table~\ref{tab1}.    

\begin{figure}[t]
  \begin{center}
	\includegraphics[width=8.6cm,height= 6.6724cm]{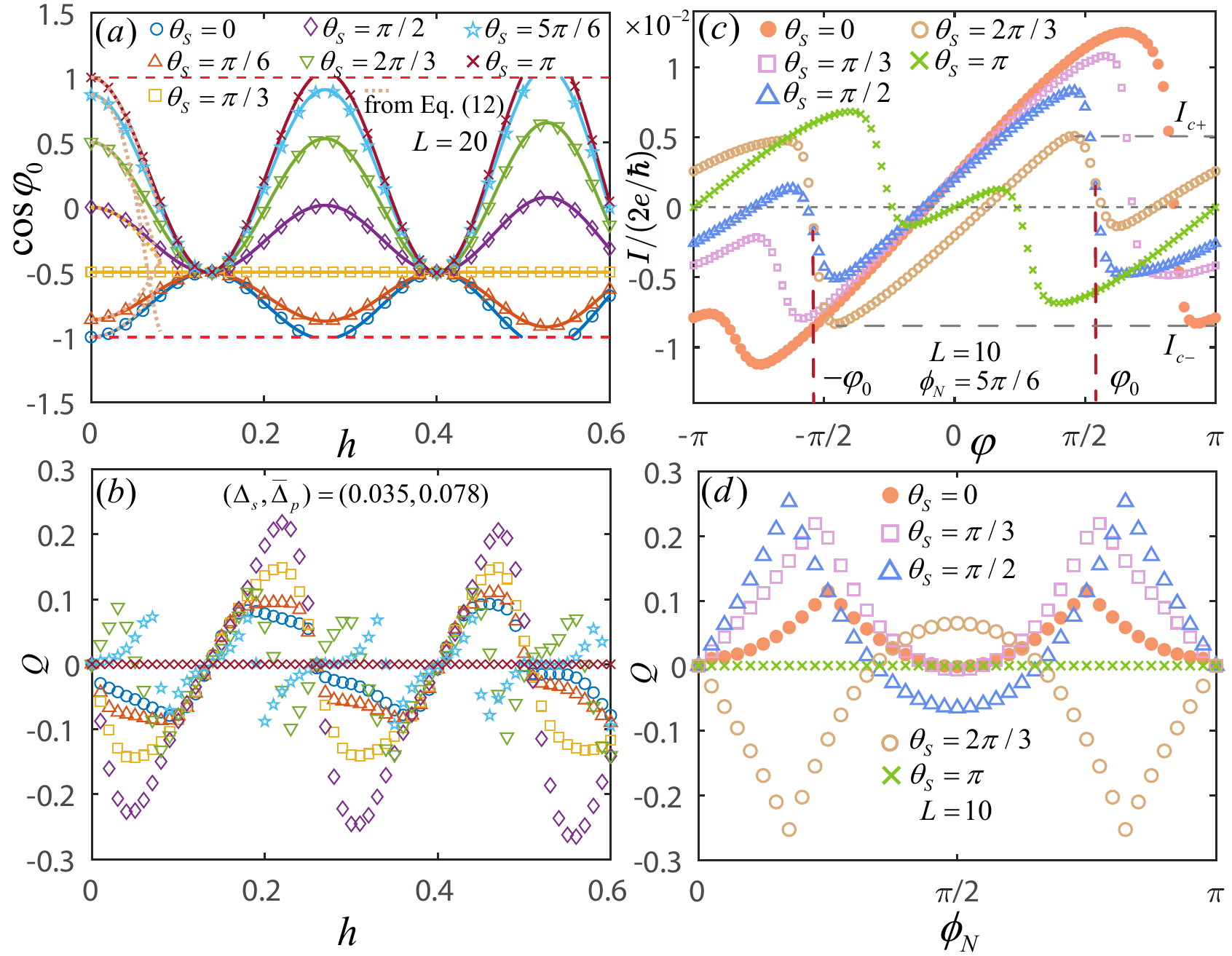}
  \end{center}
  \vspace{-0.5cm}
  \caption{ZEABSs and SDE in Josephson junction with distinct $p$-wave $\boldsymbol{d}$-vectors differing by $\theta_S$ angle. (a) Dependence of the the ZEABSs and (b) the corresponding SDE efficiency factor $Q$ as functions of the magnitude of the Zeeman field for different $\theta_S$ with $\phi_N=\pi/6$ and $k_F=\pi/3$.
(c) CPRs and (d) factor $Q$ as functions of $\phi_N$ for a $p$-wave dominant phase at different $\theta_S$, where $(\Delta_s,\Bar{\Delta}_p) =(0.035,0.045)$, $k_F=\pi/6$ and $h=0.05$. Common parameters: $t_I=-0.8$, $k_B T=4\times 10^{-4}$.}
   
\label{fig4}
\end{figure}

\section{CPR, and MF enhanced SDE with high efficiency factor}\label{IV}
It is worth noting that the results of the above symmetry analysis does not rely on whether the SC is topological or not. Both can host SDE or not. In Fig.~\ref{fig3}(a) we demonstrate centrosymmetric CPRs for several typical cases without SDE but breaking both $\boldsymbol{\mathcal{T}}$ and $\boldsymbol{\mathcal{P}}$ symmetries, including both topological nontrivial and trivial ones. With the interface roughness, however, they can be distinguished from each other by their diode efficiency factor $Q$ for the mixed $s$+$p$-wave junction hosting SDE, where $Q=\frac{I_{c+}-|I_{c-}|}{I_{c+}+|I_{c-}|}$ with $I_{c\pm}$ the current maxima along two directions. As shown in Fig.~\ref{fig3}(d), near the phase transition at $\Delta_s=\Bar{\Delta}_p$, $Q$ exhibits a great enhancement in the nontrivial regime compared to its trivial counterpart. This remarkable feature may provide an accessible signature for identifying topological phase transition through SDE measurements. Since the Josephson current contributed by ABSs is given by $I= -\frac{2e}{\hbar}\sum_{E_b>0}\frac{\partial E_b}{\partial \varphi}$, with $E_b$ the ABSs' energies, for the topological $p$-wave dominant cases, as shown in Fig.~\ref{fig3}(b), the CPRs exhibit the abrupt changes at $\pm\varphi_0$: $\Delta I=\frac{4e}{\hbar}\frac{\partial E_b}{\partial \varphi}\mid_{\varphi=\pm\varphi_0,E_b=0}$, where the exact ZEABSs reside. This gives rise to high diode efficiency factor $Q$, as shown in Fig.~\ref{fig3}(e), where as the orientation $\phi_N$ of the Zeeman field varies, $Q$ starts from a situation without SDE at $\phi_N=0$ and then form two local maxima (up to 30\%) with different signs as $\phi_N$ advances up to $\pi/2$. While for the trivial $s$-wave dominant cases, as shown in Fig.~\ref{fig3}(c), the curves exhibit smooth variations due to the absence of the ZEABSs, leading to relatively small $Q$. See Fig.~\ref{fig3}(f).

\section{Josephson junction with distinct \lowercase{\textit{d}}-vectors}\label{V}
We now focus on the situation where the SCs on the two sides of the normal link take distinct $p$-wave $\boldsymbol{d}$-vectors. This can be engineered by introducing a tiny spatial-varying magnetic field in the corresponding superconducting quantum interference device, to slowly modulate the $\boldsymbol{d}$-vector from one end to another, since $\boldsymbol{d}$ is always perpendicular to the spin quantization direction. For definiteness, we assume the two $\boldsymbol{d}$-vectors are within the $xy$ plane and differing by $\theta_S$ angle (see Fig.~\ref{fig1}). The pairing Hamiltonian then becomes: $\frac{1}{2}\sum_{k}\{\Delta_p \sin{k}(e^{i\theta^{\alpha}_S}c^{\dagger}_{k \uparrow}c^{\dagger}_{-k \uparrow}-e^{-i\theta^{\alpha}_S}c^{\dagger}_{k \downarrow}c^{\dagger}_{-k \downarrow})+\text{h.c}\}$, with $\theta^L_S=\theta_S/2$ and $\theta^R_S=-\theta_S/2$. 
Although the two $\boldsymbol d$-vectors differ, exact ZEABSs still exist. The two split level crossings are still positioned symmetrically at $\varphi=\pm\varphi_0$, protected by fermion parity. When $h=\lambda=0$, we found exactly that $\varphi_0=\pi+\theta_S$. In the presence of a small $h$, the two crossings continue to shift symmetrically as $\varphi_0=\pi +\theta_S+\Delta \varphi$, with $\Delta \varphi$ obtained analytically for a finite $\sin{\theta_S}$ (see Appendix~\ref{E}):
\begin{equation}
    \begin{aligned}
    \Delta\varphi=
    -\frac{h^2\left\{(L\sin k_F)^2-\sin^2{(k_FL)}\right\}\left(\cos{2 \phi_N}-\cos{{\theta_S}}\right)}{4\sin^4{(k_FL)}\sin{{\theta_S}}},\\  
    \end{aligned}
    \label{eq12}
\end{equation}
which remarkably exhibits the quadratic dependence on $h$, in contrast to the linear dependence for the previous $\theta_S=0$ case. When $\theta_S=2\phi_N$, the crossings are nearly locked at $\varphi_0=\pi+\theta_S$, independent of $h$. Similar to $\theta_S=0$ case, as $h$ increases, as shown in Fig.~\ref{fig4}(a), for some parameter regions of $(\theta_S, \phi_N)$, the existence region of ZEABSs show a band-like structure, while for others, the ZEABSs are always preserved and shift quasi-periodically. Interestingly, as shown in Fig.~\ref{fig4}(b), as increasing $h$, the factor $Q$ always changes sign whenever traversing the regions without ZEABSs while form local maxima at the existence regions. The robustness of these two crossings against inhomogeneity also shows similar behavior to that of $\theta_S=0$: being removed only at $\varphi_0=0$ or $\pi$ in pairs, maintaining fermion parity conserved.

To further examine the SDE, we neglect the SOC since it plays a minor role. From arguments similar to Eq.~(\ref{eq8}) and Eq.~(\ref{eq10})-(\ref{eq11}), we found that if the two $\boldsymbol{d}$-vectors are anti-parallel ($\theta_S=\pi$), or $h_x=0$ ($\phi_N=0$), the SDE is forbidden. We concentrate on a topological $p$-wave dominant junction near the phase transition. Starting from $\theta_S=\pi$ or $\phi_N=0$, states hosting SDE with relatively high $Q$ factors can be expected within $\theta_S$$\in$$(0,\pi)$ and $\phi_N$$\in$$(0,\pi/2)$. This is shown in Fig.~\ref{fig4}(c)-(d), where tunable enhancements at $\theta_S\neq\pi$ of nonreciprocal charge transport have been achieved. Thus the Zeeman field in the normal link or the tiny magnetic field in the SC can serve as switches in this kind of Josephson junction, which may turn on or off and tune the SDE by modulating $\phi_N$ or $\theta_S$. This tunability establishes a feasible way for optimizing superconducting diode performance in topological Josephson devices.  

\section*{Acknowledgments}
This work was supported by NSFC under Grant No.~11874202, and Innovation Program for Quantum Science and Technology under Grant No.2024ZD0300101.

\appendix
\renewcommand{\appendixname}{APPENDIX}
\section{JOSEPHSON CURRENT CALCULATION BY MATSUBARA GREEN'S FUNCTION}\label{A}
\begin{figure*}[t]
  \begin{center}
	\includegraphics[width=17cm,height=3.0088 cm]{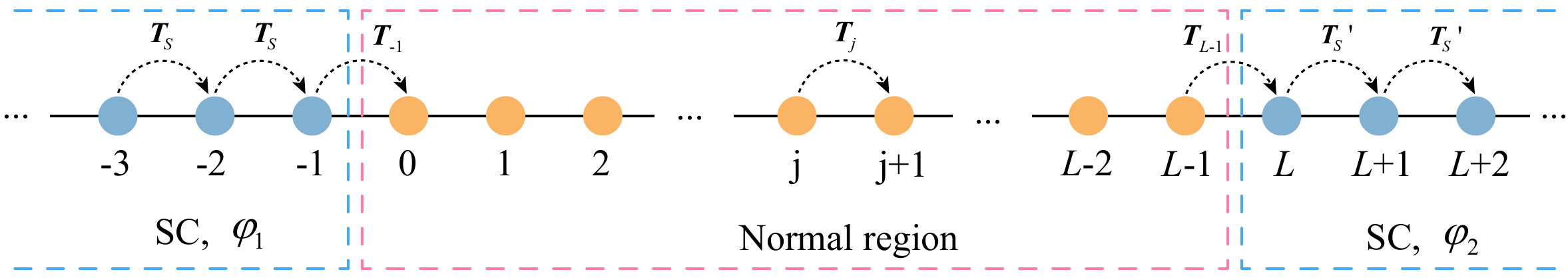}
  \end{center}
  \vspace{-0.2cm}
  \caption{One-dimensional lattice of the Josephson junction system used for current calculations in Appendix~\ref{A}-\ref{B}, where the left and right superconductors are connected by a finite-length normal metal. Both superconductors are treated as being half-infinite in Appendix~\ref{A}, but being finite with length $L_S$ in Appendix~\ref{B}. Here $\varphi=\varphi_1-\varphi_2$ is the superconducting phase difference, and $\boldsymbol T_S$ and $\boldsymbol T_S'$ are the homogeneous nearest-neighbor hopping matrices in Nambu representation for the two superconductors, while $\boldsymbol T_j$ is that from site $j$ to $j+1$ for the normal metal and interfaces.}
\label{fig5}
\end{figure*}

Here in this section we shall provide a method of calculating the Josephson current by making use of Matsubara Green's function. Physically, the charge current flowing in either superconductor of a Josephson junction consists of both supercurrent and quasi-particle current. How to calculate them separately and accurately is a rather technical problem, especially for superconductors with spin-orbit coupling or mixed pairing parity. However, the analysis can be greatly simplified by focusing on the normal link if only the total current is concerned, since, according to current conservation, the total charge current through the normal region equals that through the superconducting regions. The hybrid lattice system we consider is an infinite one, where, apart from the normal link with finite length $L$, both superconductors on the two sides of the junction are assumed to be half-infinite, as schematically shown in Fig.~\ref{fig5}.

To simplify the discussion, for the tight-binding lattice model of the normal link, it is assumed that only the nearest-neighbor hopping integrals are taken into account and each lattice site possesses no degrees of freedom other than spin. By choosing a cross section between site $j$ and $j+1$ within the normal region ($0 \leq j \leq L-1 $), the total Josephson charge current at a finite temperature can be expressed as the following thermodynamic average:   
\begin{equation}
	\begin{aligned}
		I&\equiv  I_{j+1 \gets j }\\
  &=-\frac{i e}{2\hbar} \left[ \langle \psi^{\dagger}(j+1) \tau_z  \boldsymbol{T}_{j}^{} \psi(j)    \rangle -\langle \psi^{\dagger}(j) \tau_z  \boldsymbol{T}_{j}^{\dagger} \psi(j+1)   \rangle \right]\\
        &=\frac{e}{\hbar} \text{Im}\left[ \langle \psi^{\dagger}(j+1) \tau_z  \boldsymbol{T}_{j}^{} \psi(j)       \rangle \right],\\	
	\end{aligned}
 \label{A1}
 \end{equation}
where $\tau_z$ is the Pauli matrix acting on the particle-hole space
and $\psi(j)=\left(c_\uparrow (j), c_\downarrow (j),c_\uparrow ^\dagger(j), c_\downarrow ^\dagger (j)\right)^T$ with $c_{\sigma}(j)$ denoting the annihilation operator for electrons at site $j$. Here $\boldsymbol{T}_j$ is the hopping matrix in Nambu space, 
\begin{equation}
	\begin{aligned}
    \boldsymbol{T}_j=
    \begin{pmatrix}
    {T}_j& \\
     &-{T}_j^*
    \end{pmatrix}
    \end{aligned}
     \label{A2}
 \end{equation}
 where ${T}_j$ corresponds to the nearest-neighbor hopping matrix from site $j$ to $j+1$. Introducing the standard Matsubara Green's function:
\begin{equation}
	\begin{aligned}
	\mathcal{G}_{j_1,\,j_2}(\boldsymbol \tau)&=-\langle \mathcal{T}_{\boldsymbol \tau}\psi_{j_1}(\boldsymbol \tau)\psi_{j_2}^{\dagger}(0)    \rangle\\
        &=\frac{1}{\beta} \sum_{n=-\infty }^{+\infty}   \mathcal{G}_{j_1,\,j_2}(i \omega_n) e^{-i \omega_n \boldsymbol \tau/\hbar},
	\end{aligned}
 \label{A3}
 \end{equation}
 where $\boldsymbol \tau$ is the imaginary time and $\mathcal{T}_{\tau}$ denotes the imaginary time-ordering operator, with Matsubara frequency $\omega_n=(2n+1)\pi /\beta$, $n\in{\mathrm{Z}}$, $\beta=(k_BT)^{-1}$, the average in Eq.~(\ref{A1}) becomes:
\begin{equation}
	\begin{aligned}
		&\langle \psi^{\dagger}(j+1) \tau_z  \boldsymbol{T}_{j} \psi(j)\rangle=-\text{Tr}\left\{ \tau_z  \boldsymbol{T}_{j}\langle \psi(j)  \psi^{\dagger}(j+1)\rangle\right\}\\
        &=\text{Tr}\left\{\tau_z  \boldsymbol{T}_{j} \mathcal{G}_{j,\,j+1}(\boldsymbol \tau \to 0^-)\right\}\\
        &=\frac{1}{\beta} \sum_{n=-\infty }^{+\infty} \text{Tr} \left\{ \tau_z  \boldsymbol{T}_{j} \mathcal{G}_{j,\,j+1}(i \omega_n)\right\} .
	\end{aligned}
 \label{A4}
 \end{equation}
The total charge current can then be reexpressed as:
\begin{equation}
	\begin{aligned}
        &I=\\
        &-\frac{i e}{2\hbar \beta} \sum_{n=-\infty }^{+\infty} \text{Tr} \left\{ \tau_z \left[ \boldsymbol{T}_{j} \mathcal{G}_{j,\,j+1}(i\omega_n)  -\boldsymbol{T}_{j}^{\dagger} \mathcal{G}_{j+1,\,j} (i\omega_n) \right ]\right\}.
	\end{aligned}
 \label{A5}
 \end{equation}
 In the numerical computation of the above current, the infinite sum over the Matsubara frequencies should be replaced by a finite sum of $n$ ranging from $-N$ to $N$, as long as $Nk_{B}T$ is much larger than the typical energy scale of the system, such as the gap order parameter. Then for each $\omega_n$ within the range, we compute the two Green's functions in Eq.~(\ref{A5}) by applying the recursive Green’s function method. First, we obtain the two superconductors' surface Green's functions $g_L\equiv \mathcal{G}^L_{-1,\,-1}$ and $g_R\equiv \mathcal{G}^R_{L,\,L}$ at the two interfaces between the two superconductors and the normal link. They respectively obey the following self-consistent Dyson equations:
 \begin{equation}
	\begin{aligned}
         &g_{L}^{-1}=i\omega_n\mathbf{1}_{4 \times 4}-(H_\text{BdG})_{-1,-1}-\boldsymbol{T}_{S} g_{L}\boldsymbol{T}_{S}^\dagger ,\\
         &g_{R}^{-1}=i\omega_n\mathbf{1}_{4 \times 4}-(H_\text{BdG})_{L,L}-\boldsymbol{T}_{S}'^{\dagger} g_{R} \boldsymbol{T}_{S}',\\
     \end{aligned}
      \label{A6}
 \end{equation}
 where $(H_\text{BdG})_{jj}$ is the on-site ($jj$) block matrix of the real-space BdG Hamiltonian, and $\boldsymbol {T}_S$ ($\boldsymbol {T}_S'$) is the hopping matrix in Nambu representation of the left(right)-side superconductor (see Fig.~\ref{fig5}). For simple systems, the two surface Green's functions can be obtained analytically. Otherwise, they have to be computed numerically by iteration. Secondly, starting from the known $g_L$ and $g_R$, and by a similar iteration process, one then computes the updated diagonal surface Green's functions $\mathcal{G}^L_{j,\,j}$ and $\mathcal{G}^R_{j,\,j}$ by solving the self-consistent Dyson equations recursively. Namely, in the $(j+1)$-th iteration process from the left side, we have
 \begin{equation}
        \begin{aligned}
	   \mathcal{G}_{j,\,j}^L=\left(i\omega_n \mathbf{1}_{4 \times 4}-(H_\text{BdG})_{jj}-\boldsymbol{T}_{j-1} \mathcal{G}_{j-1,\,j-1}^L \boldsymbol{T}_{j-1}^{\dagger}\right)^{-1},
       \end{aligned}
 \label{A7}
 \end{equation}
and in the $(L-j)$-th iteration process from the right side, analogously we have
 \begin{equation}
        \begin{aligned}
	   \mathcal{G}_{j,\,j}^R=\left(i\omega_n \mathbf{1}_{4 \times 4}-(H_\text{BdG})_{jj}-\boldsymbol{T}_{j}^{\dagger} \mathcal{G}_{j+1,\,j+1}^R \boldsymbol{T}_{j}^{ }\right)^{-1}.
       \end{aligned}
 \label{A8}
 \end{equation}
Now by taking the self-energy corrections from both sides into account, we have the diagonal total Green's function ${G}_{j,\,j}(i \omega_n)$ at site $j$:
\begin{equation}
        \begin{aligned}
	   \mathcal{G}_{j,\,j}(i \omega_n)
       =&\left(  i  \omega_n \mathbf{1}_{4 \times 4}-(H_\text{BdG})_{jj}\right.\\
       &\left.-\boldsymbol{T}_{j-1} \mathcal{G}_{j-1,\,j-1}^L \boldsymbol{T}_{j-1}^{\dagger}
       -\boldsymbol{T}_{j}^{\dagger} \mathcal{G}_{j+1,\,j+1}^R \boldsymbol{T}_{j}^{ }\right)^{-1}.
       \end{aligned}
 \label{A9}
 \end{equation}
Finally the needed off-diagonal total Green's functions $\mathcal{G}_{j,\,j+1}(i\omega_n)$ and $\mathcal{G}_{j+1,\,j}(i\omega_n)$ in Eq.~(\ref{A5}) can then be determined by:
\begin{equation}
        \begin{aligned}
	  \mathcal{G}_{j,\,j+1}(i\omega_n)&=\mathcal{G}_{j,\,j}^L \boldsymbol{T}_{j}^{\dagger} \mathcal{G}_{j+1,\,j+1}=\mathcal{G}_{j,\,j} \boldsymbol{T}_{j}^{ \dagger} \mathcal{G}_{j+1,\,j+1}^R,\\
        \mathcal{G}_{j+1,\,j}(i\omega_n)&=\mathcal{G}_{j+1,\,j+1} \boldsymbol{T}_{j}^{ } \mathcal{G}_{j,\,j}^L=\mathcal{G}_{j+1,\,j+1}^R \boldsymbol{T}_{j}^{ } \mathcal{G}_{j,\,j}.
       \end{aligned}
 \label{A10}
 \end{equation}
Substitution of these computation results into Eq.~(\ref{A5}) and taking the sum over the Matsubara frequencies, one finally acquire the total Josephson current flowing in the hybrid system. This framework enables efficient evaluation of the finite-temperature Josephson charge current, and is capable of computing arbitrary number of sites in the normal link \cite{SF_Green_function,GREEN1,GREEN2,GREEN3,GREEN4}, even if randomness is taken into account.

\section{JOSEPHSON CURRENT CALCULATION BY DIRECT HAMILTONIAN MATRIX DIAGONALIZATION}\label{B}
Another method of Josephson current calculation can be made if instead of considering half-infinite superconductors, the two superconductors at both sides of the normal link are treated as having finite size. Actually this method can be achieved by directly diagonalizing the whole model Hamiltonian $H_\text{BdG}$ of the hybrid junction system. In the lattice space, $H_\text{BdG}$ can be given by
\begin{equation}
    \begin{aligned}
    (H_\text{BdG})_{ij}=
    \begin{pmatrix}
    t_{ij}-(\boldsymbol{h}_i\cdot\boldsymbol{\sigma}+\mu)\delta_{ij} & 
    \mathbf\Delta_{ij}\\
    (\mathbf\Delta_{ji})^{\dagger} & -t_{ij}^*+(\boldsymbol{h}_i\cdot\boldsymbol{\sigma}+\mu)\delta_{ij}
\end{pmatrix}
    \end{aligned}
    \label{B1}
\end{equation}
where $\mu$ is the chemical potential, being identical for the whole junction system, and $t_{ij}$ is the hopping matrix from site $j$ to $i$. A possible Zeeman field $\boldsymbol{h}_i$ is also introduced, which can be different from site to site. Here we consider a general situation of pairing potential $\mathbf\Delta_{ij}$ between site $i$ and $j$, regardless of its pairing symmetry being singlet, triplet or their mixture. The real-space $\mathbf\Delta_{ij}$ is assumed to vary as follows: 
\begin{equation}
    \begin{aligned}
    \mathbf \Delta_{ij}=
    \begin{dcases}
    \Delta_{ij} e^{i\varphi_1},\;\;\;\; &-L_S\le i,j<0,\\
    \Delta_{ij} e^{i\varphi_2}, \;\;\;\;\;\;\;&L\le i,j< L+L_S,\\
    0,\;\;\;\;\;\;\;\;\;\;\;\;\;\;\;\;\;\;\;\;&\text{otherwise},
    \end{dcases}
\end{aligned}
\label{B2}
\end{equation}
where $\varphi_1$, $\varphi_2$ are the pairing phases for the superconductors, with $\varphi=\varphi_1-\varphi_2$ the superconducting phase difference. So both superconductors contain $L_S$ sites, and the whole open system has $N=2L_S+L$ lattice sites. As before, the total Josephson charge current can be computed via a cross section between site $j$ and $j+1$ in the normal link:
\begin{equation}
    \begin{aligned}
        I&= I_{j+1 \gets j }
        =\frac{2e}{\hbar}\text{Im}\{ \langle c^{\dagger}(j+1)T_j   c(j) \rangle\}\\
        &=\frac{2e}{\hbar}\sum_{\alpha,\,\beta={\uparrow,\downarrow}}\text{Im}\{(T_j)_{\alpha\beta} \langle c_{\alpha}^{\dagger}(j+1)   c_{\beta}(j) \rangle\},
    \end{aligned}
    \label{B3}
\end{equation}
 By diagonalizing the $2N\times 2N$ real-space BdG Hamiltonian for the whole junction system, one has the eigenvalue equation: $H_\text{BdG}u_n=E_n u_n $, with $u_n=[u_{n\uparrow}(1),u_{n\downarrow}(1),\cdots,u_{n\uparrow}(2N),u_{n\downarrow}(2N)]^T$ and $1\leq n\leq 2N$. Then the Bogoliubov quasi-particle excitation operators ${\gamma}_n$ can be introduced as below by an expansion of electron annihilation operator via a unitary transformation:
 \begin{equation}
c_\sigma (j)=\sum_n^{2N}{u}_{n\sigma}(j){\gamma}_n,\label{B4}
\end{equation}
where ${\gamma}_n$ satisfy the anti-commutation relations: $\{{\gamma}_n,{\gamma}_m^{\dagger}\}=\delta_{mn}$, with $1\leq n,m \leq 2N$. Therefore, we have $\left \langle {\gamma}_n^\dagger {\gamma}_m\right \rangle =f(E_n)\delta_{nm}$, and $\left \langle {\gamma}_m {\gamma}_n^\dagger \right \rangle =f(-E_n)\delta_{nm}$ with $f(E)=1/(1+e^{\beta E})$ the Fermi distribution function. But it is worth mentioning that here not all ${\gamma}_n$ are annihilation operators, nor are they independent of each other. Since, due to the intrinsic particle-hole symmetry of the BdG Hamiltonian, each eigenvalue $E_n$ must be accompanied by another eigenvalue $-E_n$, we have additionally $E_n=-E_{n-N}$ and $\gamma_{n}=\gamma_{n-N}^\dagger$ when $N<n\leq 2N$.
The thermodynamic average can then be recognized to be
\begin{equation}
    \begin{aligned}
         &\langle c_{\alpha}^{\dagger}(j+1) c_{\beta}(j) \rangle
         =\sum_{n,m=1}^{2N}{u}_{n\alpha}^*(j+1) {u}_{m\beta}(j)\left \langle {\gamma}_n^{\dagger} {\gamma}_m\right \rangle\\
         &=\sum_{n=1}^{2N}{u}_{n\alpha}^*(j+1) {u}_{n\beta}(j)f(E_n)\\
         &=-\langle c_{\beta}(j)c_{\alpha}^{\dagger}(j+1) \rangle
         =-\sum_{n=1}^{2N}{u}_{n\alpha}^*(j+1) {u}_{n\beta}(j)f(-E_n)\\
         &=-\frac{1}{2}\sum_{n=1}^{2N}{u}_{n\alpha}^*(j+1) {u}_{n\beta}(j)\tanh(\beta E_n/2).
     \end{aligned}
     \label{B5}
\end{equation}
Finally the Josephson current can be given by
\begin{equation}
    \begin{aligned}
        &I=\\
        &-\frac{e}{\hbar}\sum_{n=1}^{2N}\sum_{\alpha,\,\beta}\text{Im}\{(T_j)_{\alpha\beta}{u}_{n\alpha}^*(j+1) {u}_{n\beta}(j)\}\tanh(\beta E_n/2).
    \end{aligned}
    \label{B6}
\end{equation}

Compared to the Matsubara Green's function method for an infinite hybrid system in Appendix~\ref{A}, this method of current calculation for a finite-size hybrid system is much simpler. What's important is that it can handle the situation at the exact zero temperature. When the superconductor is topological, more attention should be paid since at each open end of the superconductors, one or more Majorana fermions will be introduced, which may cause non-negligible effect if the size is chosen too small. In actual numerical computation, as long as $L_S$ of the superconductors is sufficiently large, numerical accuracy can be controlled within a reasonable range. But the cost is that the consuming time for this method would be generally longer than that for the Green's function method. On the other hand, as a function of the superconducting phase difference $\varphi$, eigenvalues $E_n(\varphi)$ obtained here give directly the energy spectrum of the junction, and in particular, give the Andreev-bound-state (ABS) energy spectrum, as has been used in the main text. 

\section{ANALYTICAL SOLUTION OF THE ANDREEV BOUND STATES IN THE MIXED \textit{s+p}-WAVE JOSEPHSON JUNCTION WITH $h=\lambda=0$} \label{C}
Here in this section, based on the scattering matrix methods \cite{Beenakker}, we shall derive analytically the ABS energy spectrum of the Josephson junction made from a superconductor with mixed $s+p$-wave pairing. Physically, the ABS formation within the normal region involves multiple particle or hole reflections between the two S-N interfaces of the junction. To seek these states, one can invoke Eq.~(\ref{eq3}) which provides a simplified process of analyzing the ABSs by solving independently the scattering problem at each individual interface. Now we shall first find $\boldsymbol{r}_1$, $\boldsymbol{r}_2$ for the two interfaces and then combine them by the formula to derive the final analytic result. At the end of this section, a self-contained derivation of the formula will also be given.

\begin{figure*}[t]
  \begin{center}
	\includegraphics[width=17cm,height=6.8148 cm]{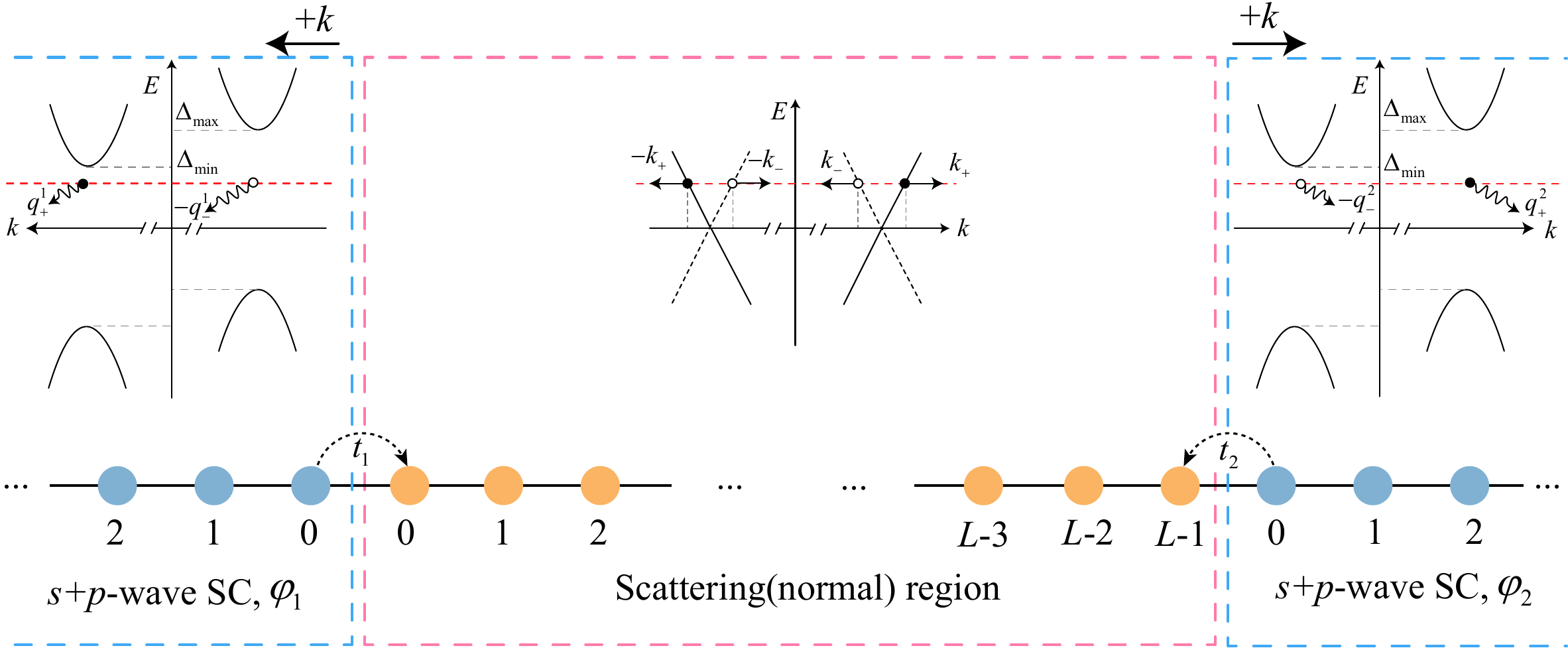}
  \end{center}
  \vspace{-0.2cm}
  \caption{One-dimensional lattice of the Josephson junction made from a superconductor with mixed $s+p$-wave pairing. Here the displayed dispersion relations at the left- and right-side superconductors illustrate the situation of (e$\uparrow$,h$\downarrow$) Nambu subspace, where the two gap minima $\Delta_{\text{max/min}}$ are given by $\Delta_{\text{max}}=\Delta_s+\Delta_p\sin k_F$, and $\Delta_{\text{min}}=\mid\Delta_s-\Delta_p\sin k_F\mid$.}
\label{fig6}
\end{figure*}

In the absence of a Zeeman field and spin-orbit interaction in the normal region, the BdG Hamiltonian describing the whole hybrid system including the superconductor and the normal link can be decoupled into two independent blocks: one is in the up-electron (e$\uparrow$) and down-hole (h$\downarrow$) space, the other is in the down-electron (e$\downarrow$) and up-hole (h$\uparrow$) space. Here the $p$-wave $\boldsymbol{d}$-vector has been chosen to be along $\bm{z}$ direction for convenience. So the problem of seeking the ABSs can be solved independently for each Nambu subspace. In either subspace, the BdG Hamiltonians in $k$-space representation for the two regions are given by:
\begin{equation}
    \begin{aligned}
    \mathcal{H}_\text{BdG}^S&=\xi_k\tau_z +(\pm\Delta_s +\Delta_p \sin{k})  \tau_x, \\
    \mathcal{H}_\text{BdG}^N&=\xi_k \tau_z,
 \end{aligned}
 \label{C1}
\end{equation}
with $\Delta_s$ in the ({e}$\uparrow$h$\downarrow$) Nambu subspace, while $-\Delta_s$ in the (e$\downarrow$h$\uparrow$) Nambu subspace. We now focus on the ({e}$\uparrow$h$\downarrow$) subspace. The quasi-particle excitation energy spectrum is $E_k=\sqrt{\xi_k^2+(\Delta_s +\Delta_p \sin{k})^2}$ and the ABSs are expected to occur when $E$ is within the gap minimum of the two superconductors, namely, $E<\Delta_{\text{min}}\equiv\mid\Delta_s-\Delta_p\sin k_F\mid$. For each interface, the incident and reflected wave functions for electron or hole incidence read below:
\begin{subequations}
\begin{equation}
    \begin{aligned}
    &\text{Left: }\Psi_N^{(1)}(j)=\\
    &\begin{cases}
        \begin{pmatrix}
        1  \\
        0
        \end{pmatrix}
        (e^{-ik_+j}+r_{ee}^{(1)}e^{ik_+j})+r_{he}^{(1)}
        \begin{pmatrix}
        0  \\
        1
        \end{pmatrix}
        e^{-ik_-j},\,\;\;\;e \text{ inc.},\\
        \begin{pmatrix}
        0  \\
        1
        \end{pmatrix}
        (e^{ik_-j}+r_{hh}^{(1)}e^{-ik_-j})+r_{eh}^{(1)}
        \begin{pmatrix}
        1  \\
        0
        \end{pmatrix}
        e^{ik_+j}, \;\;\;\;\;\,h \text{ inc.};
        \end{cases}
        \end{aligned}
        \end{equation}
        \begin{equation}
        \begin{aligned}
        &\text{Right: }\Psi_N^{(2)}(j)=\\
        &\begin{cases}
        \begin{pmatrix}
        1  \\
        0
        \end{pmatrix}
        (e^{ik_+j'}+r_{ee}^{(2)}e^{-ik_+j'})+r_{he}^{(2)}
        \begin{pmatrix}
        0  \\
        1
        \end{pmatrix}
        e^{ik_-j'},\;\;\;e \text{ inc.},\\
        \begin{pmatrix}
        0  \\
        1
        \end{pmatrix}
        (e^{-ik_-j'}+r_{hh}^{(2)}e^{ik_-j'})+r_{eh}^{(2)}
        \begin{pmatrix}
        1  \\
        0
        \end{pmatrix}
        e^{-ik_+j'},\;h \text{ inc.},
        \end{cases}
    \end{aligned}
     \label{C2}
\end{equation}
\end{subequations}
where $j'=j-(L-1)$ and $\pm k_{\pm}$ are the wave vectors of the incident or reflected waves, schematically shown in Fig.~\ref{fig6}, with the convention that $k_{\pm}>0$. The transmitted wave function in either superconductor in the corresponding scattering state is:
\begin{equation}
    \begin{aligned}
        \Psi_S^{(\alpha)}(j)
        &=t_e^{\alpha}
        \frac{1}{\sqrt{2}} 
        \begin{pmatrix}
        1  \\
        e^{i(\theta_+^{\alpha}-\varphi_\alpha)}
        \end{pmatrix}
        e^{iq_+^\alpha j}\\
        &+t_h^{\alpha}\frac{1}{\sqrt{2}}
        \begin{pmatrix}
        1  \\
        e^{-i(\theta_-^\alpha+\varphi_\alpha)}
        \end{pmatrix}
        e^{-iq_-^\alpha j},\; \alpha=1,\;2,
        \end{aligned} 
         \label{C3}
\end{equation}
where $\varphi_\alpha$ are the superconducting phases for the two superconductors and $\theta_\pm ^{\alpha}$ are defined by
\begin{equation}
\begin{aligned}
    &e^{i\theta_\pm ^1 }=\frac{E-i\sqrt{\Delta_{\mp}^{2}-E^2}}{\Delta_{\mp}},\Delta_{+}=\Delta|_{k= q_-^1},\,\Delta_{-}=\Delta|_{k=- q_+^1},\\
    &e^{i\theta_\pm ^2 }=\frac{E-i\sqrt{\Delta_{\pm}^{2}-E^2}}{\Delta_{\pm}},\, \Delta_{+}=\Delta|_{k= q_+^2},\Delta_{-}=\Delta|_{k=- q_-^2}.
    \end{aligned}
\label{C4}    
\end{equation} For a generic mixed $s+p$-wave superconductor, the four phases $\theta_\pm ^{\alpha}$ are independent for a nonzero-energy $E$. When $E=0$, according to Eq.~(\ref{C4}), $\theta_{\pm}^\alpha$ can only take values $\pm\pi/2$, although $q_{\pm}^\alpha$ and $\Delta_{\pm}$ are generally complex numbers. This feature will be used when we discuss the exact ZEABSs in Appendix~\ref{D}. At each interface, the scattering state obeys the corresponding boundary conditions as below:
\begin{subequations}
    \begin{equation}
\text{Left:}
\begin{dcases}
    \;\;\;\;\;\;\;\;\;\;\;\;\;\;\;\;\;\tau_z \Psi_N^{(1)}(-1)&=-t_1 \tau_z \Psi_S^{(1)}(0),\\
        (\tau_z-i \frac{\Delta_p}{2}\tilde\tau_1) \Psi_S^{(1)}(-1)&=-t_1 \tau_z \Psi_N^{(1)}(0);
\end{dcases}
\end{equation}\label{C5a}
\begin{equation}
\text{Right:}
\begin{dcases}
    \;\;\;\;\;\;\;\;\;\;\;\;\;\;\;\;\;\;\;\tau_z \Psi_N^{(2)}(L)&=-t_2 \tau_z \Psi^{(2)}_S(0),\\
        (\tau_z+i \frac{\Delta_p}{2} \tilde\tau_2) \Psi_S^{(2)}(-1)&=-t_2 \tau_z \Psi_N^{(2)}(L-1),
\end{dcases}
\label{C5b}
\end{equation}
\end{subequations}
where the real parameters $t_\alpha$ are the interface hopping integrals between the superconductors and the normal link and
\begin{equation}
    \begin{aligned}
    \tilde\tau_\alpha=\begin{pmatrix}
     0& e^{i\varphi_\alpha}\\
     e^{-i\varphi_\alpha} & 0\\
    \end{pmatrix},\;\alpha=1,\;2.
    \end{aligned}
    \label{C6}
\end{equation}
These boundary conditions can be derived for each interface by a combination of the Schrodinger's equations for both end sites in the normal and superconducting regions \cite{Zhou_Yao,Liu_Chun_Chi}. See Fig.~\ref{fig6}. Consider the weak-pairing limit: $\Delta_p\ll 1$, the boundary conditions can be greatly simplified by ignoring the $p$-wave pairing terms:
\begin{subequations}
\begin{equation}
\text{Left:}
\begin{dcases}
\Psi_N^{(1)}(-1)&=-t_1  \Psi_S^{(1)}(0),\;\;\;\;\;\;\;\\
\Psi_S^{(1)}(-1)&=-t_1  \Psi_N^{(1)}(0);
\end{dcases}
\end{equation}\label{C7a}
\begin{equation}
\;\;\;\;\;\text{Right:}
\begin{dcases}
 \Psi_N^{(2)}(L)&=-t_2  \Psi^{(2)}_S(0),\\
 \Psi_S^{(2)}(-1)&=-t_2  \Psi_N^{(2)}(L-1).
\end{dcases}
\end{equation}\label{C7b}
\end{subequations}
Then the reflection amplitudes are obtained by solving the above equations:
\begin{equation}
    \begin{dcases}
    r_{ee}^{(\alpha)}&=-(1-t_\alpha^2)(e^{2ik_F}-t_\alpha^2)(1-e^{i(\theta_+^\alpha + \theta_-^\alpha)})/\gamma_\alpha, \\
    r_{he}^{(\alpha)}&=4 t_\alpha^2 \sin^2 k_F e^{i( \theta_+^\alpha -\varphi_\alpha)}/\gamma_\alpha,\\
    r_{hh}^{(\alpha)}&=-(1-t_\alpha^2)(e^{-2ik_F}-t_\alpha^2)(1-e^{i(\theta_+^\alpha + \theta_-^\alpha)})/\gamma_\alpha,\\
    r_{eh}^{(\alpha)}&=4 t_\alpha^2 \sin^2 k_F e^{i( \theta_-^\alpha +\varphi_\alpha)}/\gamma_\alpha,\\
    \end{dcases}
    \label{C8}
\end{equation}
where $\gamma_\alpha=4t_\alpha^2\sin^2k_F+(1-t_\alpha^2)^2[1-e^{i(\theta_+^\alpha +\theta_-^\alpha)}]$. Here we have employed the BTK approximation $k_+\approx k_-\approx q^{\alpha}_+\approx{q^{\alpha}_-}\approx{k_F}$ \cite{BTK}, which is valid in the weak-pairing limit where the magnitude of the pairing order parameter is much less than the Fermi energy. This can be satisfied in most of the typical situations as long as the relevant energy scale is not too small. Within this approximation, we have: $\theta^1_{+}=\theta^2_{-}$ and $\theta^1_{-}=\theta^2_{+}$. But in the following discussion, we shall keep the difference between $\theta^{1}_{\pm}$ and $\theta^2_{\mp}$ to allow the possibility that the superconductors on the two sides of the link are different, having different pairing potentials $\Delta_s$ and $\Delta_p$ as an example. Now the two reflection amplitude matrices at the two interfaces are obtained, and are expressed below in a $2\times2$ standard unitary matrix form: 
\begin{equation}
    \begin{aligned}
    &\boldsymbol{r}_\alpha=\begin{pmatrix}
  r_{ee}^{(\alpha)}& r_{eh}^{(\alpha)} \\
  r_{he}^{(\alpha)}& r_{hh}^{(\alpha)}
  \end{pmatrix}=\\
  &e^{i(\frac{\theta^{\alpha}_+ +\theta^{\alpha}_-}{2}-\theta^{\alpha}_\gamma)}\begin{pmatrix}
  -e^{-i\psi_\alpha}\cos\vartheta _\alpha& e^{-i(\frac{\theta^{\alpha}_+-\theta^{\alpha}_-}{2}-\varphi_\alpha)}\sin\vartheta _\alpha \\
  e^{i(\frac{\theta^{\alpha}_+-\theta^{\alpha}_-}{2}-\varphi_\alpha)}\sin\vartheta _\alpha& e^{i\psi_\alpha}\cos\vartheta _\alpha
  \end{pmatrix}\\
  &=e^{i\boldsymbol{\varTheta}_\alpha}U(\psi_\alpha,\,\vartheta _\alpha,\,\varphi_\alpha-({\theta^{\alpha}_+-\theta^{\alpha}_-})/{2}),
    \end{aligned}
    \label{C9}
\end{equation}
where 
\begin{equation}
\begin{aligned}
&\gamma_\alpha=|\gamma_\alpha| e^{i \theta_\gamma^\alpha},\,
\boldsymbol{\varTheta}_\alpha=\frac{\theta^{\alpha}_++\theta^{\alpha}_-}{2}-\theta^\alpha_\gamma,\,\\
&\tan \psi_\alpha=\frac{\cos 2 k_F-t_\alpha^2}{\sin 2k_F},\,
\sin \vartheta _\alpha=\frac{4t_\alpha^2 \sin^2 k_F}{\left | \gamma_\alpha \right | },\,\alpha=1,2.
\end{aligned}
\label{C10}
\end{equation}
\begin{figure*}[t]
  \begin{center}
	\includegraphics[width=18cm,height=4.1287 cm]{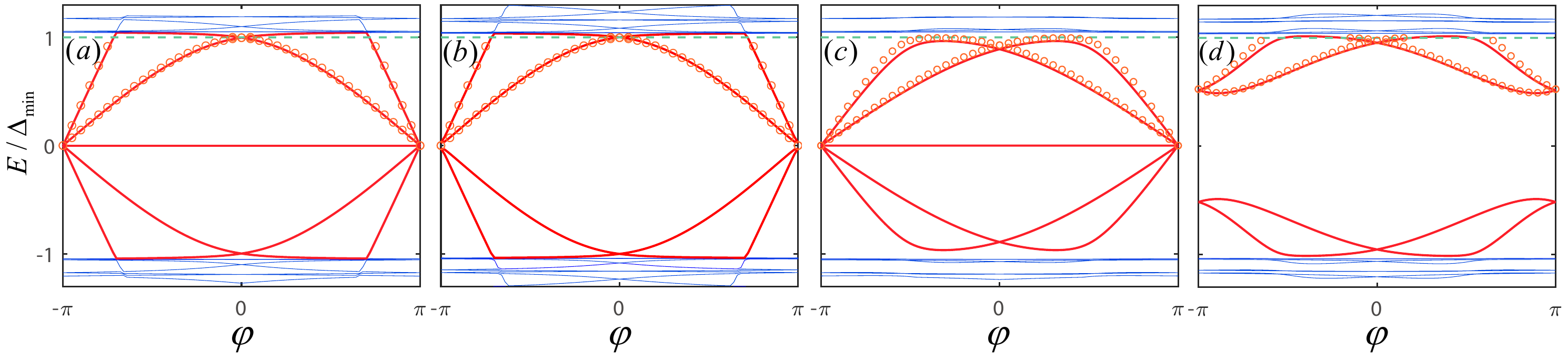}
  \end{center}
  \vspace{-0.2cm}
  \caption{Energy spectra of the ABSs in the mixed $s+p$-wave superconductor Josephson junction, as functions of the superconducting phase difference $\varphi$, where the analytical results obtained from Eq.~(\ref{C14}) (hollow circles) and the numerical results (solid lines) by Hamiltonian matrix diagonalization method introduced in Appendix~\ref{B} are compared with each other. (a), (c): $p$-wave dominant topological nontrivial cases with $(\Delta_s,\bar\Delta_p)=(0.01,0.02)$. (b), (d): $s$-wave dominant topological trivial cases with $(\Delta_s,\bar\Delta_p)=(0.02,0.01)$. Here $t_1=t_2=-1$ is shared by (a) and (b), while $t_1=t_2=-0.6$ is shared by (c) and (d). The other parameters are: $k_F=\pi/6$, $L=10$, and $L_S=1000$ in the numerical calculation.}
\label{fig7}
\end{figure*}
Notice that in the BTK approximation the propagation matrix can be given by:
\begin{equation}
    T_L=\begin{pmatrix}
  e^{ik_F (L-1)}& \\
  & e^{-ik_F (L-1)}
\end{pmatrix},
\label{C11}
\end{equation}
where the length $L$ is number of sites of the normal region. Together with the reflection matrices derived above, Eq.~(\ref{eq3}) can be expressed as:
 \begin{equation}
     \begin{aligned}
         &\text{det}\{e^{i\boldsymbol{\varTheta}_1}\times\\
         &U(\psi_1-k_F(L-1),\,\vartheta _1,\,\varphi_1-({\theta^{1}_+-\theta^{1}_-})/{2}+k_F(L-1))\\
         &-e^{-i\boldsymbol{\varTheta}_2}\times\\
         &U(-\psi_2+k_F(L-1),\,\vartheta _2,\,\varphi_2-({\theta^{2}_+-\theta^{2}_-})/{2}+k_F(L-1))\}\\
         &=0.
     \end{aligned}
     \label{C12}
 \end{equation}
 Detailed calculation shows:
\begin{equation}
    \begin{aligned}
        \cos(\boldsymbol{\varTheta}_1&+\boldsymbol{\varTheta}_2)=\cos \vartheta _1 \cos \vartheta _2 \cos(2 k_F (L-1) -\psi_1-\psi_2)\\
        &\;\;\;\;+\sin \vartheta _1 \sin \vartheta _2 \cos\left(\varphi-\frac{\theta^{1}_+ - \theta^{1}_-}{2} +\frac{\theta^{2}_+ - \theta^{2}_-}{2}\right),
    \end{aligned}
    \label{C13}
\end{equation}
After some simplifications, we acquire the central result of this section: the analytic equation determining the ABS spectrum $E_b(\varphi)$,
\begin{equation}
\begin{aligned}
    &\cos \frac{1}{2}(\theta^{1}_+ + \theta^{1}_- + \theta^{2}_+ + \theta^{2}_-)\\
    -&\cos\left(\varphi-\frac{\theta^{1}_+ - \theta^{1}_-}{2} +\frac{\theta^{2}_+ - \theta^{2}_-}{2} \right)
    \\
    =&2\sin{\frac{\theta^{1}_++\theta^{1}_-}{2}}\sin{\frac{\theta^{2}_++\theta^{2}_-}{2}} \left\{ \frac{(1-t_1^2)^2}{(\hbar v_F)^2 t_1^2}+\frac{(1-t_2^2)^2}{(\hbar 
 v_F)^2t_2^2}\right.\\
&+\frac{2(1-t_1^2)^2(1-t_2^2)^2}{(\hbar  v_F)^4 t_1^2 t_2^2}
+\frac{2(1-t_1^2)(1-t_2^2)}{(\hbar  v_F)^4 t_1^2 t_2^2}\\
&[-\cos 2k_F(L+1)+ (t_1^2+t_2^2)\cos 2k_F L\\
 &\left.-t_1^2 t_2^2\cos 2k_F(L-1) ]\right\} ,
\end{aligned}
\label{C14}
\end{equation}
with $\hbar v_F=2\sin k_F$. It can be easily shown that an equation similar to Eq.~(\ref{C14}) can be derived for the (e$\downarrow$h$\uparrow$) subspace, just with the replacement of $\varphi$ by $-\varphi$. For identical superconductors with $\theta^1_{+}=\theta^2_{-}\equiv\theta_{-}$ and $\theta^1_{-}=\theta^2_{+}\equiv\theta_{+}$, we recover Eq.~(\ref{eq4}) in the main text. In Fig.~\ref{fig7}, we compare this analytical result with the corresponding numerical one calculated by matrix diagonalization method introduced in Appendix~\ref{B}. They agree with each other very well.  

Analogously, we can derive the analytic result of the ABSs for the corresponding continuum model. In this continuum model, as in the BTK theory \cite{BTK}, the boundary conditions for the left interface ($x=0$) and those for the right interface ($x=L$) should be:
\begin{subequations}
\begin{equation}
\text{Left:}
\begin{dcases}
    \Psi_N(0)=\Psi_S(0),\\
    \left.\frac{d \Psi_S(x)}{dx} \right|_{x=0}-\left.\frac{d \Psi_N(x)}{dx} \right| _{x=0}=-2k_F Z_1 \Psi_S(0),
\end{dcases}
\end{equation}\label{C15a}
\begin{equation}
\text{Right:}
\begin{dcases}
    \Psi_N(L)=\Psi_S(L),\\
    \left.\frac{d \Psi_S(x)}{dx}\right| _{x=L}-\left.\frac{d \Psi_N(x)}{dx}\right|_{x=L}=2k_F Z_2 \Psi_S(L),
\end{dcases}
\end{equation}\label{C15b}
\end{subequations}
where the dimensionless parameters $Z_\alpha=U_\alpha/\hbar v_F$ ($\alpha=1,2$) denote the effective strength of the barrier potential $U_1\delta(x)$ and $U_2\delta(x-L)$ at the interfaces. By an exactly same computation process, we obtain the corresponding analytic equation determining the ABSs in the case of the continuum model:
\begin{equation}
\begin{aligned}
    &\cos \frac{1}{2}(\theta^{1}_+ + \theta^{1}_- + \theta^{2}_+ + \theta^{2}_-)\\
    -&\cos\left(\varphi-\frac{\theta^{1}_+ - \theta^{1}_-}{2} +\frac{\theta^{2}_+ - \theta^{2}_-}{2} \right)\\
    =&2\sin{\frac{\theta^{1}_++\theta^{1}_-}{2}}\sin{\frac{\theta^{2}_++\theta^{2}_-}{2}}\{Z_1^2+ Z_2^2+ 2Z_1^2Z_2^2\\
    +&2Z_1 Z_2 [(1-Z_1 Z_2) \cos 2k_FL+(Z_1+Z_2)\sin 2k_FL)]\}.
\end{aligned}
\label{C16}
\end{equation}
Based on this equation we now demonstrate for identical superconductors some particular cases, where $Z_1=Z_2=Z/2$ and $L=0$ are also chosen for simplicity. For the purely $s$-wave case ($\Delta_p=0$), $\Delta_-=\Delta_+=\Delta_s$, with $\theta_{+}=\theta_{-}$, independent of $k$, Eq.~(\ref{C16}) can be simplified to be $\cos{2\theta_+}-\cos{\varphi}=2 Z^2 \sin^2{\theta_+}$. According to the relation $E=\Delta_s \cos{\theta_+}$, one can find the energy $E_b$ of the ABSs satisfies $E_b= \Delta_s \sqrt{1-\sin^2{(\varphi/2)}/(1+Z^2)}$. As long as $Z\neq 0$, there exists no $E_b=0$ ABS. While for the purely $p$-wave case ($\Delta_s=0$), we have $\Delta_+=-\Delta_-$, resulting in $\theta_{+}+\pi=\theta_{-}$. Eq.~(\ref{C16}) can be simplified to $-\cos{2\theta_+}+\cos{\varphi}=2 Z^2 \cos^2{\theta_+}$. Its solution leads to the energy of the ABSs: $E_b=\Delta_p \sin{k_F}\cos{(\varphi/2)}/\sqrt{1+Z^2}$. Thus a pair of exact ZEABSs always exists, namely, $E_b=0$ is always a solution when $\varphi=\pi$ no matter how large $Z$ is.

Now we give a self-contained derivation of Eq.~(\ref{eq3}) as the condition for generic bound states existing within a homogeneous intermediate region between two interfaces. If there are $2m$ propagating modes within the region: $m$ from right to left, and $m$ from left to right, let us denote $\Psi_i^{(L)}$($\Psi_i^{(R)}$) as the scattering states with the former (latter) $m$ propagating modes as the incident waves. So $\Psi_i^{(L)}$($\Psi_i^{(R)}$) exactly match the boundary conditions at the left (right) interface, and we further denote respectively the corresponding $m\times m$ reflection matrices as $\boldsymbol{r}_1$ and $\boldsymbol{r}_2$. Any bound state between the interfaces is assumed to be the superposition of all the $2m$ scattering states:
\begin{equation}
    \begin{aligned}
        \Psi_{\text{Bound}}=\sum_{i=1}^m\{a_i \Psi_i^{(L)} +b_i \Psi_i^{(R)}\},
    \end{aligned}
    \label{C17}
\end{equation}
where $a_i$ and $b_i$ are the superposition coefficients. The reasonability of this state requires that the first series of terms have to match the boundary conditions at the right interface, while the second series of terms have to match the boundary conditions at the left interface. Therefore, the incident modes on the left interface must come from the superposition of the reflection modes from the right interface, and vice versa. So one must require:
\begin{equation}
\begin{aligned}
    T_L \boldsymbol{r}_{1} a=b,\;T_L \boldsymbol{r}_{2} b=a,
\end{aligned}
\label{C18}
\end{equation}
where $T_L=\text{diag}\{e^{i k_1 (L-1)},e^{i k_2 (L-1)} ,\cdots,e^{i k_m (L-1)}\}$ represents the propagation of the propagating modes between the interfaces, and $a=(a_1,a_2,\cdots,a_m)^T$, $b=(b_1,b_2,\cdots,b_m)^T$. Combination of the two equations leads to:
\begin{equation}
\begin{aligned}
    (T_L \boldsymbol{r}_{2} T_L \boldsymbol{r}_{1}-1)a=0,\;(T_L \boldsymbol{r}_{1} T_L \boldsymbol{r}_{2}-1)b=0.
    \end{aligned}
    \label{C19}
\end{equation}
Because at least one of $a$ or $b$ should be nonzero, the determinant of the coefficient matrices should be zero. This leads to Eq.~(\ref{eq3}) or its equivalent statements on the conditions of the existence of the bound states. 

\section{EXACT ZEABSS IN A TOPOLOGICAL SUPERCONDUCTOR JOSEPHSON JUNCTION: $\theta_S=0$ CASE}\label{D}

\begin{figure*}[t]
  \begin{center}
	\includegraphics[width=17cm,height=4.04405 cm]{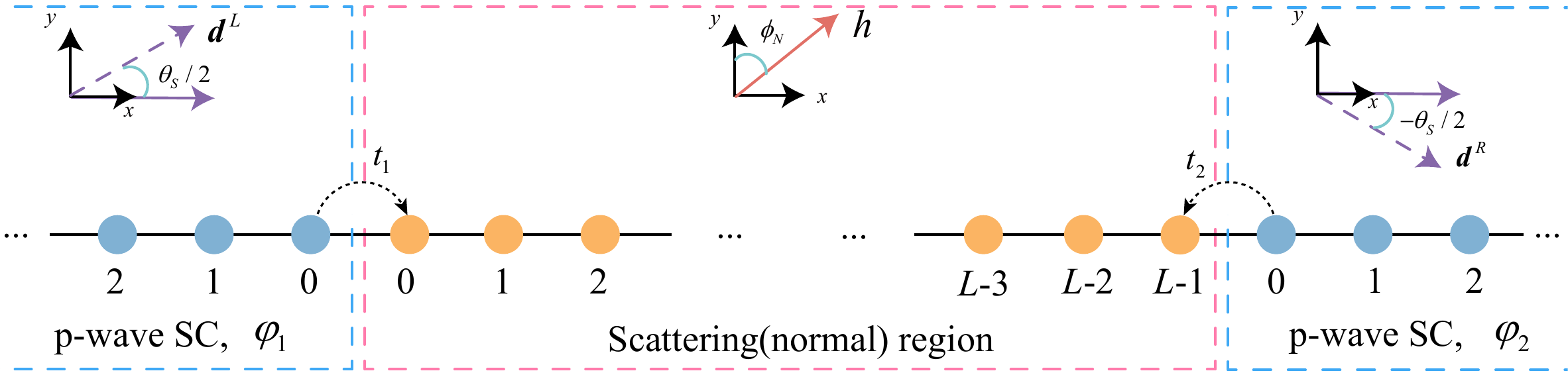}
  \end{center}
  \vspace{-0.2cm}
  \caption{One-dimensional lattice of the Josephson junction made from the mixed $s+p$-wave superconductors, where $\boldsymbol{d}^L$ and $\boldsymbol{d}^R$ are the $\boldsymbol{d}$-vectors of the $p$-wave pairing in the superconductors on both sides, which may be identical (Appendix~\ref{D}) or be differing by $\theta_s$ angle (Appendix~\ref{E}). A Zeeman field $\boldsymbol{h}$ within the $xy$ plane is also applied in the normal link.}
\label{fig8}
\end{figure*}

The significance of the superconducting diode effect in Josephson junctions critically depends on the ZEABSs and their phase-dependent characteristics. Generally a trivial superconductor Josephson junction cannot guarantee the existence of the ZEABSs. In this and next sections, for a Josephson junction composed by a topological superconductor, we shall exhibit in detail the existence of the exact ZEABSs and their stability within a finite range of parameters. Our conclusions can be established for a generic $p$-wave dominant mixed $s+p$-wave superconductor, but for simplicity, we shall analyze as an example a purely $p$-wave superconductor. To get the information about the ZEABSs, we will still make use of the single-interface-coupling method of Eq.~(\ref{eq3}). The Hamiltonians for the superconducting and normal regions can be given by:
\begin{equation}
    \begin{aligned}
    \mathcal{H}_S&=\xi_k\tau_z +\Delta_p \sin{k}  \sigma_z \tau_x,\\
    \mathcal{H}_N&=\xi_k \tau_z+h \sin{\phi_N}\sigma_x\tau_z+h\cos{  \phi_N} \sigma_y,
 \end{aligned}
 \label{D1}
\end{equation}
where without loss of generality the $\boldsymbol{d}$-vector of the $p$-wave superconductor is assumed to be along $\boldsymbol{x}$, and a Zeeman field $\boldsymbol{h}$ within $xy$ plane is applied in the normal link. In the following, we shall always fix energy to be $E=0$ to seek the ZEABSs, and to further ensure their exactness, we shall also do the derivation without any approximation. In the final step, we will acquire an equation. When there exists a solution to this equation, an exact ZEABS also do exist. Otherwise, there are no ZEABSs. In the following discussion of this section, for convenience, we always denote the spin parallel or opposite to the Zeeman field as $\Uparrow$ or $\Downarrow$, respectively.

First, we consider the special case of $\phi_N=\pi/2$, namely, the Zeeman field being along the same direction to the $\boldsymbol{d}$-vector: $\boldsymbol{h}=h(1,\,0,\,0)$. For a spin-$\Uparrow$ electron incident on the left interface from the normal region, the incident and the reflected wave function reads:
\begin{equation}
    \begin{aligned}
        &\Psi_N(j)=\\
        &\frac{1}{\sqrt{2}}\begin{pmatrix}
        1  \\
        1\\
        0\\
        0\\
        \end{pmatrix}
        (e^{-ik_F^+ j}+r_{e}^\Uparrow e^{ik_F^+ j})
        +\frac{1}{\sqrt{2}}\begin{pmatrix}
        1  \\
        -1\\
        0\\
        0\\
        \end{pmatrix}
        r_{e}^\Downarrow e^{ik_F^- j}\\
        &+
        \frac{1}{\sqrt{2}}\begin{pmatrix}
        0  \\
        0\\
        1\\
        1\\
        \end{pmatrix}
        r_{h}^\Uparrow e^{-ik_F^+ j}
        +\frac{1}{\sqrt{2}}\begin{pmatrix}
        0  \\
        0\\
        1\\
        -1\\
        \end{pmatrix}
        r_{h}^\Downarrow e^{-ik_F^- j},\\
    \end{aligned}
    \label{D2}
\end{equation}
where $k_F^{\pm}$ are the Fermi wave vectors for the spin-$\Uparrow$ and spin-$\Downarrow$ electrons in the normal region. Accordingly the transmitted wave function in the left-side superconductor reads
\begin{equation}
    \begin{aligned}
        &\Psi_S^{(1)}(j)=\\
        &\frac{1}{\sqrt{2}}\left\{ \begin{pmatrix}
        1  \\
        0\\
        e^{i(\theta^{1}_+-\varphi_{1})}\\
        0\\
        \end{pmatrix}
        t_e^{\uparrow}e^{iq_+j}+
        \begin{pmatrix}
        1  \\
        0\\
        -e^{i(\theta^{1}_- -\varphi_{1})}\\
        0\\
        \end{pmatrix}
        t_h^\uparrow e^{-iq_-j}\right.\\
        &\left.+\begin{pmatrix}
        0  \\
        1\\
        0\\
        e^{i(\theta^{1}_--\varphi_{1})}\\
        \end{pmatrix}
        t_{e}^\downarrow e^{iq_+ j}+
        +\begin{pmatrix}
        0  \\
        1\\
        0\\
        -e^{i(\theta^{1}_+-\varphi_{1})}\\
        \end{pmatrix}
        t_{h}^\downarrow e^{-iq_- j}\right\}\\
        &=\frac{1}{\sqrt{2}}\begin{pmatrix}
        1  \\
        0\\
        i e^{-i\varphi_{1}}\\
        0\\
        \end{pmatrix}
        (t_e^{\uparrow}e^{iq_+j}+t_h^\uparrow e^{-iq_-j})\\
        &+\frac{1}{\sqrt{2}}\begin{pmatrix}
        0  \\
        1\\
        0\\
        -i e^{-i\varphi_{1}}\\
        \end{pmatrix}
        (t_{e}^\downarrow e^{iq_+ j}+t_{h}^\downarrow e^{-iq_- j}),\\
    \end{aligned}
    \label{D3}
\end{equation}
which includes the electron-like ($q_+$) and hole-like ($-q_-$) transmitted waves for each spin index. Here ${\theta_{\pm}^1}=\pm \pi/2$ have been used, which gives rise to the coalesce of the eigenvectors in pairs in the above wave function. 

Invoking the boundary conditions similar to Eq.~(\ref{C5a}), we have
\begin{equation}
\begin{dcases}
    \;\;\;\;\;\;\;\;\tau_z \Psi_N(-1)&=-t_1 \tau_z \Psi_S^{(1)}(0),\\
    (\tau_z-i\frac{\Delta_p}{2}\sigma_z\tilde \tau_1) \Psi_S^{(1)}(-1)&=-t_1 \tau_z \Psi_N(0).
\end{dcases}
\label{D4}
\end{equation}
For the exactness of the derivation, $k_F^+$ and $k_F^-$ with small difference between them are distinguished and the reflection amplitudes from spin-$\Uparrow$ to spin-$\Downarrow$ then should be renormalized by the group velocities (multiplied by an additional factor $\sqrt{\sin k_F^-/\sin k_F^+}$). The final result of the reflection amplitudes for this electron-$\Uparrow$ incidence is:
\begin{equation}
    \begin{dcases}
    r_{ee}^{ \Uparrow \Uparrow}&=e^{ik_F^+}\frac{\tan{(k_{F}^+/2)}-\tan{(k_{F}^-/2)}}{\tan{(k_{F}^+/2)}+\tan{(k_{F}^-/2)}}\equiv e^{ik_F^+}\sin{ \Theta }, \\
    r_{ee}^{ \Downarrow \Uparrow}&=0,\\
    r_{he}^{ \Uparrow \Uparrow}&=0,\\
    r_{he}^{ \Downarrow \Uparrow}&=i e^{i(\theta^{\alpha}_+-\varphi_{\alpha}+\frac{\Delta k_F}{2})}\frac{2\sqrt{\tan{(k_{F}^+/2)}\tan{(k_{F}^-/2)}}}{\tan{(k_{F}^+/2)}+\tan{(k_{F}^-/2)}}\\
    &\equiv i e^{i(\theta^{\alpha}_+-\varphi_{\alpha}+\frac{\Delta k_F}{2})}\cos{ \Theta },\\
    \end{dcases}
    \label{D5}
\end{equation}
where $\theta_+^1=-\theta_+^2=\pi/2$ and $\Delta k_F=k_F^+-k_F^-$. It can be seen that these reflection amplitudes are independent of the parameters $\Delta_p$, $t_1$ and $q_{\pm}$, although the boundary conditions are relevant to these parameters. Actually because the two coalesced eigenvectors in Eq.~(\ref{D3}) are also the eigenvectors of $i\sigma_z\tau_z\tilde \tau_1$ with the same eigenvalue $-1$, the wave functions satisfy $\Psi_N(-1)=-t_1\Psi_S(0)$ and $\Psi_N(0)=-t_1^{-1}(1+\Delta_p/2)\Psi_S^1(-1)$. Therefore, only the four combined parameters $t_1(t_e^{\uparrow}+t_h^\uparrow)$, $t_1(t_{e}^\downarrow+t_{h}^\downarrow)$, $t_1^{-1}(1+\Delta_p/2)(t_e^{\uparrow}e^{-iq_+}+t_h^\uparrow e^{iq_-})$ and $t_1^{-1}(1+\Delta_p/2)(t_{e}^\downarrow e^{-iq_+}+t_{h}^\downarrow e^{iq_-})$ are entered as a whole into the boundary conditions, eliminating the effect of the detailed values of $\Delta_p$, $t_1$, $q_{\pm}$ on the reflection amplitudes needed. This is the technical reason for the robustness of the $E=0$ ABSs if they exist.

\begin{figure*}[t]
  \begin{center}
	\includegraphics[width=17.5cm,height=6.7536 cm]{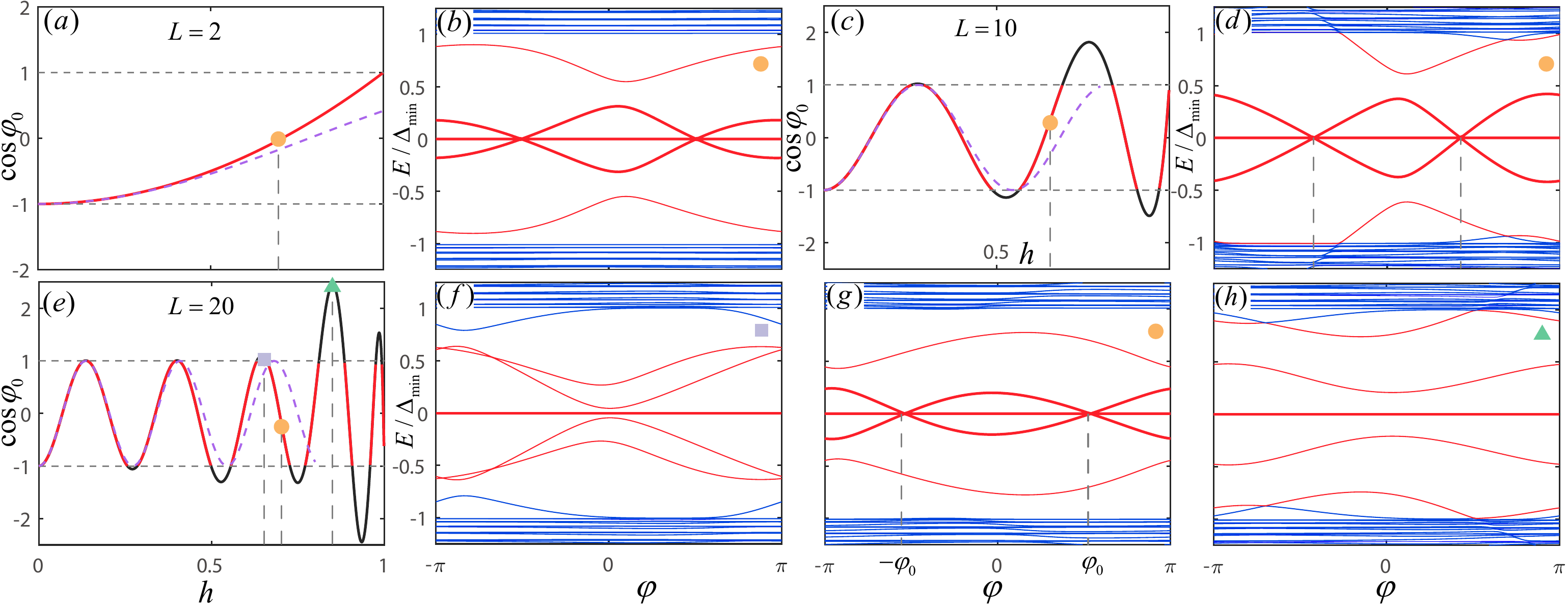}
  \end{center}
  \vspace{-0.2cm}
  \caption{Existence of the exact ZEABSs in a topological superconductor Josephson junction, in comparison with the numerical results. The exact analytic results are shown in (a), (c) and (e) with different $L$, where the expression on the right of Eq.~(\ref{D10}) as a function of the magnitude of the Zeeman field is plotted. The solid curves exhibit the oscillating behavior, giving rise to the band structure of the region of the ZEABSs. The dashed lines are the other analytic results based on Eq.~(\ref{50}), which fit the corresponding solid curves quite well at small $h$. Here the solid circles denote the exact zero-energy solutions of the ABSs, while the other symbols indicate the absence of the ZEABS. The numerical results of the energy spectra for the ABSs relevant to these five representative states are shown respectively in (b), (d), and (f)-(h). Here a mixed $s+p$-wave superconductor is considered and the parameters are chosen as: $(\Delta_s,\bar\Delta_p)=(0.02,0.104)$, $k_F=\pi/3$, $t_1=t_2=-0.6$ and $L_S=1000$.}
\label{fig9}
\end{figure*}

After considering all the four incident waves (e$\Uparrow$, e$\Downarrow$, h$\Uparrow$, h$\Downarrow$) on the left interface, the reflection amplitude matrix can be given by  
\begin{widetext}
\begin{equation}
    \begin{aligned}
    \boldsymbol{r}_{\alpha}&=\begin{pmatrix}
        r_{ee}^{ \Uparrow \Uparrow} & r_{ee}^{ \Uparrow \Downarrow}&r_{eh}^{ \Uparrow \Uparrow} &r_{eh}^{ \Uparrow \Downarrow} \\
    r_{ee}^{ \Downarrow \Uparrow}& r_{ee}^{ \Downarrow \Downarrow}&r_{eh}^{ \Downarrow \Uparrow} & r_{eh}^{ \Downarrow \Downarrow} \\
    r_{he}^{ \Uparrow \Uparrow} & r_{he}^{ \Uparrow \Downarrow}& r_{hh}^{ \Uparrow \Uparrow}& r_{hh}^{ \Uparrow \Downarrow}\\
    r_{he}^{ \Downarrow \Uparrow} &r_{he}^{ \Downarrow \Downarrow} &r_{hh}^{ \Downarrow \Uparrow} &r_{hh}^{ \Downarrow \Downarrow} \\
    \end{pmatrix}\\
    &=\begin{pmatrix}
    e^{i k_F^+}\sin{ \Theta } &0&0 &-i e^{-i(\theta^{\alpha}_+-\varphi_{\alpha})+i\frac{\Delta k_F}{2}}\cos{ \Theta }\\
    0 & -e^{i k_F^-}\sin{ \Theta }&-i e^{-i(\theta^{\alpha}_+-\varphi_{\alpha})-i\frac{\Delta k_F}{2}}\cos{ \Theta } & 0\\
      0 &i e^{i(\theta^{\alpha}_+-\varphi_{\alpha})-i\frac{\Delta k_F}{2}}\cos{ \Theta }& e^{-i k_F^+}\sin{ \Theta } & 0\\
       i e^{i(\theta^{\alpha}_+-\varphi_{\alpha})+i\frac{\Delta k_F}{2}}\cos{ \Theta } &0&0 &  -e^{-i k_F^-} \sin{ \Theta }
    \end{pmatrix}.
    \end{aligned}
    \label{D6}
\end{equation}
\end{widetext}
Obviously, this matrix can be divided into two independent submatrices, indicating that the possible ABSs can be decoupled into (e$\Uparrow$h$\Downarrow$) and (e$\Downarrow$h$\Uparrow$) branches. Therefore, we only need to calculate separately the two $2\times2$ unitary reflection matrices $\tilde{r}_\alpha$ and its complex conjugate $\tilde{r}_\alpha^*$, with $\tilde{r}_\alpha$ given by:
\begin{equation}
    \begin{aligned}
    \tilde{r}_\alpha=\begin{pmatrix}
         e^{i k_F^+} \sin{ \Theta } &  -i e^{-i(\theta^{\alpha}_+-\varphi_{\alpha})+i\frac{\Delta k_F}{2}}\cos{ \Theta }\\
        i e^{i(\theta^{\alpha}_+-\varphi_{\alpha})+i\frac{\Delta k_F}{2}}\cos{ \Theta } & - e^{-i k_F^-}\sin{ \Theta }
    \end{pmatrix}
    \end{aligned}
    \label{D7}
\end{equation}
where $k_F\equiv(k_F^+ + k_F^-)/2$. Notice that here the corresponding propagation matrices also become $\tilde{T}_L$ and its complex conjugate $\tilde{T}_L^*$, with $\tilde{T}_L$ given by,
\begin{equation}
\begin{aligned}
    \tilde{T}_L=\begin{pmatrix}
        e^{ik_{F}^+(L-1)} & \\
         & e^{-ik_{F}^-(L-1)}
    \end{pmatrix}.
    \end{aligned}
    \label{D8}
\end{equation}
Subsequently, according to Eq.~(\ref{eq3}), the fourth-order determinant can be factorized into the following second-order determinant and its complex conjugate:   \begin{widetext}
\begin{equation}
    \begin{aligned}
   \text{det}\left( \tilde T_L \tilde r_1 -\tilde r_2^\dagger \tilde T_L^\dagger\right)=&\text{det}
    \left\{\begin{pmatrix}
        \sin{ \Theta } e^{i k_F^+ L} & -i e^{-i(\theta_{+}^1-\varphi_1-k_F^+ (L-1)-\frac{\Delta k_F}{2})} \cos{ \Theta }\\
        i e^{i(\theta_{+}^1-\varphi_1-k_F^- (L-1)+\frac{\Delta k_F}{2})} \cos{ \Theta } & - \sin{ \Theta } e^{-i k_F^- L}
    \end{pmatrix}\right.\\
    -&\left.\begin{pmatrix}
        \sin{ \Theta } e^{-i k_F^+ L} & -i e^{-i(\theta_{+}^2-\varphi_2-k_F^- (L-1)+\frac{\Delta k_F}{2})} \cos{ \Theta }\\
        i e^{i(\theta_{+}^2-\varphi_2-k_F^+ (L-1)-\frac{\Delta k_F}{2})} \cos{ \Theta } & - \sin{ \Theta } e^{i k_F^- L}
    \end{pmatrix}\right\}\\
    =&4\text{det}\begin{pmatrix}
     i\sin{ \Theta } \sin{k_F^+L} & - e^{i(\tilde{\varphi}+ k_F (L-1))}\cos{ \Theta }\cos{\frac{\varphi+\Delta k_F L}{2}}\\
    - e^{-i(\tilde{\varphi}+ k_F (L-1))}\cos{ \Theta }\cos{\frac{\varphi-\Delta k_F L}{2}} &  i\sin{ \Theta } \sin{k_F^-L}
     \end{pmatrix},
    \end{aligned}
    \label{D9}
\end{equation}
\end{widetext}
where $\tilde{\varphi}=(\varphi_1+\varphi_2)/2$. For both submatrices, the equations are identical to each other. So if $E=0$ level crossings exist, they must come in pairs, and obey the following exact equation:
\begin{equation}
    \begin{aligned}
    \cos{ \varphi}=-\cos{\Delta k_F L}+\tan^2{ \Theta }(\cos{2 k_F L}-\cos{\Delta k_F L}),
    \end{aligned}
    \label{D10}
\end{equation}
where $\varphi=\varphi_1-\varphi_2$. In the absence of the Zeeman field, $\Delta k_F=0$, $\Theta=0$, one has a doubly degenerate $E=0$ level crossings at $ \varphi_0= \pi$. In the presence of a Zeeman field within the $xy$ plane, the degenerate crossings will split and their positions are expected to deviate from $\pi$, as long as $h$ is not too large, which is shown in Fig.~\ref{fig9}(a), (c) and (e). When $h$ is sufficiently large, the above equation can not ensure a real solution, indicating the absence of the ZEABSs, as shown in Fig.~\ref{fig9}(e), (f) and (h). However, the right side of the above equation exhibits an oscillating behavior as a function of $h$, resulting in the region of the ZEABSs occurring alternatively, no matter how large $h$ is. Because the equation is exact, this band-like structure of the ZEABSs is also exact. As long as $h$ is sufficiently small, the existence of the ZEABSs is always guaranteed. Now we consider in detail the limiting case of $h\ll 1$, where one can adopt the approximation: $\tan{ \Theta }\approx \sin{ \Theta } \approx  \Theta$, $\cos\Delta k_F L\approx1-(\Delta k_F L)^2/2,\,\cos{ \varphi}\approx -1+(\Delta \varphi)^2/2$. After some simplifications, we find a pair of zero-energy level crossings located at $\pm\varphi_0$, with $\varphi_0=\pi+\Delta \varphi$ and the phase shift $\Delta \varphi$ obeying: 
\begin{equation}
    \begin{aligned}
    (\Delta \varphi)^2&=(\Delta k_FL)^2-4\sin^2{ \Theta }\sin^2{k_FL},
    \end{aligned}
    \label{D11}
\end{equation}
where
\begin{equation}
    \begin{aligned}
        \sin{ \Theta}&\equiv \frac{\tan{(k_{F}^+/2)}-\tan{(k_{F}^-/2)}}{\tan{(k_{F}^+/2)}+\tan{(k_{F}^-/2)}}\approx\frac{\Delta k_F}{4\tan{\frac{k_F}{2}}\cos^2{\frac{k_F}{2}}}\\
        &=\frac{\Delta k_F}{2\sin k_F}\approx \frac{h}{2\sin^2{k_F}}.
    \end{aligned}
    \label{D12}
\end{equation}
Since $\Delta k_F\approx2h/\hbar v_F=h/\sin{k_F}$, the phase shift $\Delta \varphi$ of the ZEABSs is found to be
\begin{equation}
    \begin{aligned}
  \Delta \varphi&=\frac{h}{\sin^2{k_F}}\sqrt{(L\sin k_F)^2-\sin^2{k_FL}}.
    \end{aligned}
    \label{50}
\end{equation}
The expression is meaningful since the quantity in the square root is positive-definite when $k_F\in(0,\pi)$. This indicates that the shift of the superconducting phase difference $\Delta \varphi$ is proportional to the magnitude of the Zeeman field and is also sensitive to the length of the normal link.

In the situation of the continuum model, we can make a discussion along a similar way. Instead of Eq.~(\ref{D5}), the refection amplitudes for an electron-$\Uparrow$ incidence should be replaced by:
\begin{equation}
    \begin{dcases}
    r_{ee}^{ \Uparrow \Uparrow}&=\frac{k_{F}^+-k_{F}^-}{k_{F}^++k_{F}^-}\equiv \sin{ \Theta }, \\
    r_{ee}^{ \Downarrow \Uparrow}&=0,\\
    r_{he}^{ \Uparrow \Uparrow}&=0,\\
    r_{he}^{ \Downarrow \Uparrow}&=i e^{i(\theta^{\alpha}_+-\varphi_{\alpha})}\frac{\sqrt{k_{F}^+k_{F}^-}}{k_{F}^++k_{F}^-}
    \equiv i e^{i(\theta^{\alpha}_+-\varphi_{\alpha})}\cos{ \Theta }.\\
    \end{dcases}
    \label{D14}
\end{equation}
Compared with Eq.~(\ref{D5}), we find that besides the replacement from $\tan{(k_F^{\pm}/2)}$ to $k_F^{\pm}/2$, the phases $e^{ik_F^+}$ and $e^{i\Delta k_F/2}$ are also absent. This is because in the lattice model, the boundary conditions connect the two regions' wave functions differing by a lattice site, while here in the continuum model, the boundary conditions connect the two regions' wave functions at the same location. For the ZEABSs, along the same process, we arrive at the same conclusion to Eq.~(\ref{D10}) and Eq.~(\ref{D11}), except that here $\Delta k_F=2h/\hbar v_F\approx h k_F/E_F $ with $v_F=\hbar k_F/m$, $E_F=\hbar^2k_F^2/2m$, and $\sin{\Theta}$ in Eq.~(\ref{D12}) should be replaced by 
\begin{equation}
    \begin{aligned}
    \end{aligned}
        \sin{\Theta}\equiv \frac{{k_{F}^+}-{k_{F}^-}}{k_{F}^+ +k_{F}^-}=\frac{\Delta k_F}{2k_F}
        \approx \frac{h}{2E_F}.
    \label{D15}
\end{equation}
Then instead of Eq.~(\ref{50}), in the continuum model the phase shift $\Delta \varphi$ of the ZEABSs can be derived as:
\begin{equation}
    \begin{aligned}
     \Delta \varphi&=\frac{h}{E_F}\sqrt{(k_F L)^2-\sin^2{k_FL}}.
    \end{aligned}
    \label{D16}
\end{equation}

When the Zeeman field is applied along $\boldsymbol{z}$ direction, namely, $\boldsymbol{h}=h(0,\,0,\,1)$, the system can be divided into two independent subsystems. However, the Zeeman field only changes the chemical potential of the two subsystems. As a result, $\Delta k_F=0$ is still satisfied for each subsystem, insuring that $ \varphi_0= \pi$ still holds.

Next, for a generic direction of the Zeeman field in the $xy$ plane with $\bm{h}=h(\sin{  \phi_N},\cos{  \phi_N},0)$, we now try to solve the problem first in the framework of the continuum model and then in the lattice one, since the corresponding intermediate expressions and steps are more simpler in the former case. In the continuum model, the incident and reflected wave function in the normal region for an electron-$\Uparrow$ incidence on the left interface can be written as
\begin{widetext}
\begin{equation}
    \begin{aligned}
        \Psi_N^{e\Uparrow}(x)&=\frac{1}{\sqrt{2}}\left\{
        \begin{pmatrix}
    1 \\
        i e^{-i\phi_N}\\
        0\\
        0\\
        \end{pmatrix}
        (e^{-ik_F^+ x}+r_{e}^\Uparrow e^{ik_F^+ x})\right.
        &+\begin{pmatrix}
        1  \\
        -i e^{-i\phi_N}\\
        0\\
        0\\
        \end{pmatrix}
        r_{e}^\Downarrow e^{ik_F^- x}
        &\left.+\begin{pmatrix}
        0  \\
        0\\
        1\\
        -i e^{i\phi_N}\\
        \end{pmatrix}
        r_{h}^\Uparrow e^{-ik_F^+ x}
        +\begin{pmatrix}
        0  \\
        0\\
        1\\
        i e^{i\phi_N}\\
        \end{pmatrix}
        r_{h}^\Downarrow e^{-ik_F^- x}\right\}\\
    \end{aligned}
    \label{D17}
\end{equation}
\end{widetext}
while the transmitted wave function in the superconductor remains nearly unchanged, as given by Eq.~(\ref{D3}), with the replacement of $j$ by $x$. Using the boundary conditions in Eq.~(\ref{C15a}), the reflection amplitudes are obtained as
\begin{widetext}
\begin{equation}
    \begin{dcases}
    r_{ee}^{ \Uparrow \Uparrow}&=\frac{({k_{F}^{+}}^2-{k_{F}^{-}}^2)\sin^2{  \phi_N}}{4k_{F}^{+}k_{F}^{-}+\sin^2{  \phi_N}(k_{F}^{+}-k_{F}^{-})^2}\equiv r_A, \\
    r_{ee}^{ \Downarrow \Uparrow}&=i\frac{2\sqrt{k_{F}^{+}k_{F}^{-}}({k_{F}^{+}}-{k_{F}^{-}})\sin{ \phi_N}\cos{ \phi_N}}{4k_{F}^{+}k_{F}^{-}+\sin^2{ \phi_N}(k_{F}^{+}-k_{F}^{-})^2}\equiv i r_B,\\
    r_{he}^{ \Uparrow \Uparrow}&=i \frac{4 k_{F}^{+}k_{F}^{-}\cos{ \phi_N}e^{i(\theta_+^{(\alpha)}-\varphi_\alpha-\phi_N)}}{4k_{F}^{+}k_{F}^{-}+\sin^2{ \phi_N}(k_{F}^{+}-k_{F}^{-})^2}
    \equiv i e^{i(\theta_+^{(\alpha)}-\varphi_\alpha-\phi_N)} r_C,\\
    r_{he}^{ \Downarrow \Uparrow}&=-\frac{2\sqrt{k_{F}^{+}k_{F}^{-}}({k_{F}^{+}}+{k_{F}^{-}})\sin{ \phi_N}e^{i(\theta_+^{(\alpha)}-\varphi_\alpha-\phi_N)}}{4k_{F}^{+}k_{F}^{-}+\sin^2{ \phi_N}(k_{F}^{+}-k_{F}^{-})^2}
    \equiv -e^{i(\theta_+^{(\alpha)}-\varphi_\alpha-\phi_N)}r_D.\\
    \end{dcases}
    \label{D18}
\end{equation}
\end{widetext}
By a similar process arriving at Eq.~(\ref{D6}), the $4\times 4$ unitary reflection amplitude matrices for the two interfaces can be obtained as
\begin{widetext}
\begin{equation}
    \begin{aligned}
    \boldsymbol{r}_{\alpha}&=\begin{pmatrix}
        r_A & -i r_B&-i e^{-i(\theta_+^{(\alpha)}-\varphi_\alpha-\phi_N)}r_C  &- e^{-i(\theta_+^{(\alpha)}-\varphi_\alpha-\phi_N)}r_D \\
    i r_B& -r_{A} &- e^{-i(\theta_+^{(\alpha)}-\varphi_\alpha-\phi_N)} r_{D} & -i e^{-i(\theta_+^{(\alpha)}-\varphi_\alpha-\phi_N)}r_{C} \\
    i e^{i(\theta_+^{(\alpha)}-\varphi_\alpha-\phi_N)}r_{C} & - e^{i(\theta_+^{(\alpha)}-\varphi_\alpha-\phi_N)}r_{D}& r_{A}& i r_{B}\\
    - e^{i(\theta_+^{(\alpha)}-\varphi_\alpha-\phi_N)}r_{D} &i e^{i(\theta_+^{(\alpha)}-\varphi_\alpha-\phi_N)}r_{C} &-i r_{B} &-r_{A} \\
    \end{pmatrix}=\boldsymbol{r}_{\alpha}^{\dagger}.
    \end{aligned}
    \label{D19}
    \end{equation}
\end{widetext}
Then we are led to a fourth-order determinant that cannot be block-diagonalized. Since the deviation of the phase difference $ \Delta \varphi$ is expected to be still proportional to the Zeeman field, to facilitate the calculation, let $\Delta \varphi=\frac{h}{E_F}k_FL\zeta$. Then the polynomial equation satisfied by the dimensionless parameter $\zeta$ can be written as 
\begin{widetext}
\begin{equation}
    \begin{aligned}
    &\begin{vmatrix} 
    b\sin^2{ \phi_N} &-i e^{i\Delta k_F L/2}b\sin{ \phi_N}\cos{ \phi_N}&i e^{i(\widetilde{\phi}+k_F^+L)}\cos{ \phi_N} \zeta& e^{i(\widetilde{\phi}+k_FL)}\sin{ \phi_N}(\zeta-1)\\
    i e^{-i\Delta k_F L/2}b\sin{ \phi_N}\cos{ \phi_N}&-b\sin^2{ \phi_N}&  e^{i(\widetilde{\phi}+k_FL)}\sin{ \phi_N}(\zeta+1)&i e^{i(\widetilde{\phi}+k_F^-L)}\cos{ \phi_N} \zeta\\
    i e^{-i(\widetilde{\phi}+k_F^+L)}\cos{ \phi_N} \zeta&-e^{-i(\widetilde{\phi}+k_FL)}\sin{ \phi_N}(\zeta-1)&
    - b\sin^2{ \phi_N}&-i e^{-i\Delta k_F L/2}b\sin{ \phi_N}\cos{ \phi_N}\\
    - e^{-i(\widetilde{\phi}+k_FL)}\sin{ \phi_N}(\zeta+1)&i e^{-i(\widetilde{\phi}+k_F^-L)}\cos{ \phi_N} \zeta&
    i e^{i\Delta k_F L/2}b\sin{ \phi_N}\cos{ \phi_N}&b\sin^2{ \phi_N}\\
    \end{vmatrix}\\
    =
     &\begin{vmatrix} 
    b\sin^2{ \phi_N} &-b\sin{ \phi_N}\cos{ \phi_N}&-\cos{ \phi_N} \zeta& -\sin{ \phi_N}(\zeta-1)\\
    -b\sin{ \phi_N}\cos{ \phi_N}&-b\sin^2{ \phi_N}& -\sin{ \phi_N}(\zeta+1)&\cos{ \phi_N} \zeta\\
    \cos{ \phi_N} \zeta&\sin{ \phi_N}(\zeta-1)&
    -b\sin^2{ \phi_N}&b\sin{ \phi_N}\cos{ \phi_N}\\
    \sin{ \phi_N}(\zeta+1)&-\cos{ \phi_N} \zeta&
    b\sin{ \phi_N}\cos{ \phi_N}&b\sin^2{ \phi_N}\\
    \end{vmatrix}=0,
     \end{aligned}
     \label{D20}
\end{equation}
\end{widetext}
where $b=\sin{k_FL}/k_FL$ and $\widetilde{\phi}=\phi_N+(\varphi_1+\varphi_2)/2$. The final equation is
\begin{equation}
    \begin{aligned}
    \zeta^4-2(1-b^2)\sin^2{ \phi_N}\zeta^2+\sin^4{ \phi_N}(1-b^2)^2=0,
     \end{aligned}
     \label{D21}
\end{equation}
where the left is a perfect-square polynomial of $\zeta^2$. We have the repeated root: $\zeta=\sin{ \phi_N}\sqrt{1-b^2}$. Thus we have a pair of zero-energy level crossings located at $\pm\varphi_0=\pi\pm\Delta \varphi$ with the deviation of the phase difference $\Delta \varphi$ given by 
\begin{equation}
    \begin{aligned}
     \Delta \varphi&=\frac{h}{E_F}\sin{ \phi_N}\sqrt{(k_F L)^2-\sin^2{k_FL}},
    \end{aligned}
    \label{D22}
\end{equation}
which is consistent with Eq.~(\ref{D16}). For the lattice model, with the explanation of the difference shown in Eq.~(\ref{D12}) and Eq.~(\ref{D15}), we have the corresponding result:
\begin{equation}
    \begin{aligned}
     \Delta \varphi&=\frac{h}{\sin^2{k_F}}\sin{ \phi_N}\sqrt{(L\sin{k_F})^2-\sin^2{k_FL}}.
    \end{aligned}
    \label{D23}
\end{equation}

In addition, we also discover that besides the Zeeman field, the Rashba spin-orbit interaction will also induce the location shift of $\varphi_0$ away from $\pi$. We consider the following Hamiltonian in the normal region
\begin{equation}
    \begin{aligned}
    \mathcal{H}_N(k)=\xi_k \tau_z+\lambda \sin{k} \sigma_z.
 \end{aligned}
 \label{D24}
\end{equation}
It can be easily seen that the (e$\uparrow$h$\uparrow$) subspace is still decoupled with (e$\downarrow$h$\downarrow$) one in seeking the exact $E=0$ ABSs. We shall still carry out the derivation without any approximation. Take the former subspace as an example, and for an electron e$\uparrow$ incidence on the left interface:
\begin{equation}
    \begin{aligned}
    \Psi_N(j)=
    \begin{pmatrix}
        1  \\
        0
        \end{pmatrix}
        (e^{ -ik_F^-j}+r_{e}e^{ik_F^+ j})+r_{h}
        \begin{pmatrix}
        0  \\
        1
        \end{pmatrix}
        e^{-i k_F^+ j},
 \end{aligned}
 \label{D25}
\end{equation}
where the Fermi wave vectors $k_F^{\pm}$ obey $-2\cos{k_F^+}-\mu+\lambda\sin{k_F^+}=-2\cos{k_F^-}-\mu-\lambda\sin{k_F^-}=0$, with the convention that $k_F^{\pm}>0$. Notice that the relevant three Fermi velocities can be checked to be identical so there is no need to renormalize the reflection amplitudes. The transmitted wave function in both superconductors always read at $E=0$:
\begin{equation}
    \begin{aligned}
        &\Psi_S^{(\alpha)}(j)
        =
        \frac{1}{\sqrt{2}} 
        \begin{pmatrix}
        1  \\
         e^{i(\theta_+^{\alpha}-\varphi_\alpha)}
        \end{pmatrix}
        \left(t_e^{\alpha}e^{iq_+^\alpha j}+t_h^{\alpha} e^{-iq_-^\alpha j}\right),\; %\alpha=1,\;2.
        \end{aligned} 
        \label{D26}
\end{equation}
Invoking the boundary conditions, which are modified by the Rashba spin-orbit interaction, we have:
\begin{equation}
    \begin{aligned} 
    \begin{dcases}
    \;\;\;\;\;\;\;\;(\tau_z-i\frac{\lambda}{2}) \Psi_N(-1)&=-t_1 \tau_z \Psi_S^{(1)}(0),\\
    (\tau_z-i\frac{\Delta_p}{2}\tilde \tau_1) \Psi_S^{(1)}(-1)&=-t_1 \tau_z \Psi_N(0).
    \end{dcases}
    \end{aligned} 
    \label{D27}
\end{equation}
By eliminating the transmission amplitudes together with the $\Delta_p$ term, as discussed previously in this section, we obtain the following equations involving only the reflection amplitudes:
\begin{equation}
    \begin{aligned}
    \begin{dcases}
    &e^{i(\theta_+^1-\varphi_1)}\left(1-i \frac{\lambda}{2}\right)(e^{i k_F^-}+r_e e^{-i k_F^+})\\
    &\;\;\;\;\;\;\;\;\;\;\;\;\;\;\;\;\;\;\;\;\;\;\;\;\;\;\,\,=\left(1+i \frac{\lambda}{2}\right)r_h e^{i k_F^+},\\
    &e^{i(\theta_+^1-\varphi_1)}(1+r_e)=r_h.
    \end{dcases}
    \end{aligned} 
    \label{D28}
\end{equation}
Since $k_F^+$ and $-k_F^-$ satisfies the equation $(1+i\lambda/2)\beta^2+\mu\beta+(1-i\lambda/2)=0$ with $\beta=e^{ik}$, $e^{i (k_F^+ -k_F^-)}=\beta_1\beta_2=(1-i\lambda/2)/(1+i\lambda/2)$. So one finds that $r_e=0$, $r_h=e^{i(\theta_+^1-\varphi_1)}$ is the solution of the above equations. This means the total Andreev reflection occurs for this incidence at $E=0$. Most importantly, this solution is independent of the detailed parameters such as the magnitudes of $\Delta_p$, $t_1$ and $\lambda$, indicating the robustness of this solution. Analogously, the other amplitudes can be obtained, and the reflection amplitude matrices for the two interfaces are finally given by 

\begin{equation}
    \begin{aligned}
    r_\alpha=\begin{pmatrix}
        0 & e^{-i(\theta_+^\alpha-\varphi_\alpha)}\\
        e^{i(\theta_+^\alpha-\varphi_\alpha)} &0
    \end{pmatrix}.
    \end{aligned} 
    \label{D29}
\end{equation}
Then according to Eq.~(\ref{eq3}), we have 
\begin{equation}
    \begin{aligned}
    \text{det}(T_1 r_1 T_2 r_2-1)=0\Rightarrow\text{det}(T_1 r_1-  r_2^\dagger T_2^\dagger)=0,
    \end{aligned} 
    \label{D30}
\end{equation}
where the propagation matrices between the interfaces are different and given respectively by $T_1=\text{diag}\{e^{ik_F^+(L-1)},\;e^{-ik_F^+ (L-1)}\}$ and $T_2=\text{diag}\{e^{ik_F^-(L-1)},\;e^{-ik_F^- (L-1)}\}$. We obtain the exact result: $(\cos{(\varphi+\Delta k_F (L-1))}+1)^2=0$, indicating a doubly degenerate solution: $\varphi_0=\pi-\Delta k_F (L-1)$, required by the particle-hole symmetry. A parallel discussion of the other (e$\downarrow$h$\downarrow$) subspace results in another exact doubly degenerate solution: $\varphi_0=\pi+\Delta k_F (L-1)$. So in the presence of a Rashba spin-orbit interaction, the locations of the superconducting phase differences for the exact ZEABSs have been shifted to be $\pm\varphi_0$ with:
\begin{equation}
    \begin{aligned}
    \varphi_0=\pi- \Delta k_F (L-1),
 \end{aligned}
 \label{D31}
\end{equation}
where in the small-coupling limit of $\lambda\ll 1$, $\varphi_0$ becomes $\varphi_0=\pi+ \lambda(L-1)$ since $\Delta k_F=k_F^+-k_F^-\approx-\lambda$. 

When $\boldsymbol{d}$-vector is parallel to the spin-orbit coupling, namely, $\boldsymbol{d}_k$ is along $\bm{z}$, we find the ZEABSs are always fixed at $\varphi_0=\pi$, which can be proved similarly. Here the (e$\uparrow$h$\downarrow$) subspace is decoupled with (e$\downarrow$h$\uparrow$) one we take the former subspace as an example. For an electron e$\uparrow$ incidence on the left interface:
\begin{equation}
    \begin{aligned}
    \Psi_N(j)=
    \begin{pmatrix}
        1  \\
        0
        \end{pmatrix}
        (e^{ -ik_F^-j}+r_{e}e^{ik_F^+ j})+r_{h}
        \begin{pmatrix}
        0  \\
        1
        \end{pmatrix}
        e^{-i k_F^- j},
 \end{aligned}
 \label{D32}
\end{equation}
The transmitted wave function in both superconductors are also described by Eq.~(\ref{D26}). The boundary conditions in the (e$\uparrow$h$\downarrow$) subspace is:
\begin{equation}
    \begin{aligned} 
    \begin{dcases}
    \;\;\;\;\;\;(1-i\frac{\lambda}{2})\tau_z \Psi_N(-1)&=-t_1\tau_z  \Psi_S^{(1)}(0),\\
    (\tau_z-i\frac{\Delta_p}{2}\tilde \tau_1) \Psi_S^{(1)}(-1)&=-t_1 \tau_z \Psi_N(0).
    \end{dcases}
    \end{aligned} 
    \label{D33}
\end{equation}
Similarly we obtain the following equations involving only the reflection amplitudes:
\begin{equation}
    \begin{aligned}
    \begin{dcases}
    e^{i(\theta_+^1-\varphi_1)}(e^{i k_F^-}+r_e e^{-i k_F^+})=r_h e^{i k_F^-},\\
    e^{i(\theta_+^1-\varphi_1)}(1+r_e)=r_h.
    \end{dcases}
    \end{aligned} 
    \label{D34}
\end{equation}
The solution is also given by Eq.~(\ref{D29}). But the propagation matrices between the interfaces are replaced by
$T_1=\text{diag}\{e^{ik_F^+(L-1)},\;e^{-ik_F^- (L-1)}\}$ and $T_2=\text{diag}\{e^{ik_F^-(L-1)},\;e^{-ik_F^+ (L-1)}\}$ in Eq.~(\ref{D30}). Finally we obtain the exact result:  $(\cos{\varphi}+1)^2=0$, indicating a doubly degenerate level crossings at $\varphi_0=\pi$ since a parallel discussion and same conclusion can be made for the other (e$\downarrow$h$\uparrow$) subspace.

In the next section we shall explore the situation where the superconductors at the two sides of the link possess distinct $\boldsymbol{d}$-vectors. But one exceptional case is when they are opposite to each other, which will be considered now. For example, if $\boldsymbol{d}_k$ is along $\boldsymbol{x}$ in the left superconductor, while along $-\boldsymbol{x}$ in the right superconductor, the minus sign of $\boldsymbol{d}_k$ in the right superconductor can actually be absorbed by the pairing phase $\varphi_2$. By this absorption, this situation is just equivalent to the one where both superconductors share the same $\boldsymbol{d}_k$ along $\boldsymbol{x}$, with a superconducting phase difference $\varphi+\pi$. So in the presence of the Zeeman field, the locations $\pm\varphi_0$ of the two symmetrical zero-energy level crossings should be:
\begin{equation}
    \begin{aligned}
    \varphi_0=
    \begin{dcases}
        \frac{h}{E_F}\sin{ \phi_N}\sqrt{(k_F L)^2-\sin^2{k_FL}},\;\;\;\;\;\;
        \text{continuum,}\\
        \frac{h}{\sin^2{k_F}}\sin{ \phi_N}\sqrt{(L\sin{k_F})^2-\sin^2{k_FL}},\,
        \text{lattice}.
    \end{dcases}
    \end{aligned}
    \label{D35}
\end{equation}

Now we give a brief discussion on the $p$-wave dominant mixed $s+p$-wave pairing case, as has been examined in the main text. The central results obtained in this section for the purely $p$-wave case still hold true for the mixed pairing case until $\Delta_s$ is increased up to $\Delta_p\sin k_F$ where the superconducting gap is closed and the phase transition happens. Actually, for an $s$-wave dominant mixed $s+p$-wave Josephson junction, we have ${\theta_{+}^1}={\theta_{-}^1}=\pi/2$ and the four eigenvectors in the transmitted wave function in the main text are always kept distinct from each other. For general parameters of $\Delta_s$, $\Delta_p$, $t_1$ and $t_2$, the existence of the ZEABSs cannot be expected.   

\section{EXACT ZEABSS IN A JOSEPHSON JUNCTION COMPOSED OF TOPOLOGICAL SUPERCONDUCTORS: ${\theta_S}\neq 0$}\label{E}
In this section we further consider the situation of the Josephson junction composed by two mixed $s+p$-wave superconductors with distinct $\boldsymbol{d}$-vectors differing by an angle ${\theta_S}$, to elaborate the discussion in main text. We shall present the analytical results for the robust ZEABSs, in particular their location shifts of the superconducting phase differences as functions of the Zeeman field applied within the normal link. For simplicity, we only consider the case of the purely $p$-wave pairing. The final result does not change if instead the $p$-wave dominant mixed $s+p$-wave pairing is considered. The pairing Hamiltonians take the form
 \begin{equation}
    \begin{aligned}
    H_{p\text{-wave}}^{(\alpha)}=&
    \sum_{k}\left(\Delta_p \sin{k} e^{i(\varphi_\alpha+\theta_S^\alpha)}c^{\dagger}_{k \uparrow}c^{\dagger}_{-k \uparrow}\right.\\
    &\left.-\Delta_p \sin{k} e^{i(\varphi_\alpha-{\theta}_S^\alpha)}c^{\dagger}_{k \downarrow}c^{\dagger}_{-k \downarrow}+\text{h.c.}\right),
    \end{aligned}
    \label{E1}
\end{equation}
where ${\theta}_S^\alpha$($\alpha=1,2$) are the angles between the $\boldsymbol{d}$-vectors of the superconductors and the $\boldsymbol{x}$ axis, with both $\boldsymbol{d}$-vectors being within the $xy$ plane. Without loss of generality we take $ \theta_S^1={\theta_S}/2$, and $ \theta_S^2=-{\theta_S}/2$, where ${\theta_S} \in (0,\pi)$. See Fig.~\ref{fig8}. Now the two $\boldsymbol{d}_k$-vectors can be expressed as: $\boldsymbol{d}_k=-\Delta_p \sin{k}(\cos{\frac{\theta_S}{2}},\pm\sin{\frac{\theta_S}{2}},0)$. For the scattering state at either interface, the transmitted wave function in either superconductor can be expressed as:
\begin{equation}
    \begin{aligned}
        \Psi^{\alpha}_s(x)&=
        \frac{1}{\sqrt{2}}\begin{pmatrix}
        1  \\
        0\\
        (-1)^{\alpha+1} ie^{-i({\varphi_{\alpha}+\theta}_s^\alpha)}\\
        0\\
        \end{pmatrix}
        (t_e^{\uparrow}e^{iq_+x}+t_h^\uparrow e^{-iq_-x})\\
        &+\frac{1}{\sqrt{2}}\begin{pmatrix}
        0  \\
        1\\
        0\\
        (-1)^{\alpha} ie^{-i(\varphi_{\alpha}-{\theta}_s^\alpha )}\\
        \end{pmatrix}
        (t_{e}^\downarrow e^{iq_+ x}+t_{h}^\downarrow e^{-iq_- x}).
    \end{aligned}
    \label{E2}
\end{equation}
In the following we further assume $\varphi_1=\varphi/2$, and $\varphi_2=-\varphi/2$, and adopt the continuum model just for simplicity. When we obtain the final central result we will also give the corresponding result in the lattice version.

First, we consider the case absence of a Zeeman field in the normal region, i.e., $h=0$. In the process of seeking the ABSs, the two subspace of (e$\uparrow$h$\uparrow$) and (e$\downarrow$h$\downarrow$) are still decoupled, and the effect of $\theta_S$ is just shifting the pairing phase of $\uparrow\uparrow$ pairing by $\theta_S^\alpha$, while that of $\downarrow\downarrow$ pairing by $-\theta_S^\alpha$. So the superconducting phase difference $\varphi$ is effectively shifted by $\pm\theta_S$, the locations $\pm\varphi_0$ of the two zero-energy level crossings should be $\varphi_0=\pi +{\theta_S}$.
Then we investigate the effect of a finite $h$ on these locations. As before, the Zeeman field is treated as a small quantity relative to the Fermi energy, so the shifts of these locations relative to $\pi\pm{\theta_S}$ are also expected to be small. For the sake of concreteness, a Zeeman field is assumed to be applied within the $xy$ plane in the normal region, namely, $\bm{h}=h(\sin{ \phi_N},\,\cos{ \phi_N},\,0)$, and we consider the shift of the zero-energy level crossing at $\pi +{\theta_S}$, i.e., $\varphi_0=\pi +{\theta_S}+\Delta\varphi$. In the following we shall derive in detail the analytical correction $\Delta\varphi$.  Here the determinant of Eq.~(\ref{eq3}) can be shown to be given by
\begin{widetext}
\begin{equation}
    \begin{aligned}
    &\begin{vmatrix}
    A_+ &-i e^{i\Delta k_F L/2}B&- e^{i(k_F^+ L+\phi_N)}C& i e^{i(k_F L+\phi_N)}D_+\\
    i e^{-i\Delta k_F L/2}B&-A_-& i e^{i(k_F L+\phi_N)}D_-&- e^{i(k_F^- L+\phi_N)}C\\
    - e^{-i(k_F^+ L+\phi_N)}C^*&-i e^{-i(k_F L+\phi_N)}D_+^*&A_+^*& i e^{-i\Delta k_F L/2}B^*\\
    -i e^{-i(k_F L+\phi_N)}D_-^*&- e^{-i(k_F^- L+\phi_N)}C^*&-i e^{i\Delta k_F L/2}B^*&-A_-^*\\
    \end{vmatrix}=\begin{vmatrix}
    \,\,\,\,A_+ &-B&\,\,\,\,C& \,\,\,\,D_+\\
    -B&-A_-& \,\,\,\,D_-&-C\\
    \,\,\,\,C^*&\,\,\,\,D_+^*&\,\,\,\,A_+^*&-B^*\\
    \,\,\,\,D_-^*&-C^*&-B^*&-A_-^*\\
    \end{vmatrix}\\
    =&\left|A_+A_-+B^2\right|^2+\left|C^2+D_+ D_-\right|^2-|B|^2\left(|D_+|^2+|D_-|^2-2|C|^2\right)
    -(|A_+|^2+|A_-|^2)|C|^2\\
    &+2 \text{Re}\{A_+ A_-^{*} D_- D_+^*\}
    +2 \text{Re}\left\{\left(A_+ B^*+A_-^*B\right)\left(C D_+^{*}+C^*D_-\right)\right\}=0,
     \end{aligned}
     \label{E3}
\end{equation}
\end{widetext}
where $k_F^{\pm}=k_F\pm\Delta k_F/2$ and the matrix entries
\begin{equation}
    \begin{aligned}
    &A_{\pm}=A_1 e^{ik_{F}^{\pm}L}-A_2 e^{-ik_{F}^{\pm}L},\\
    &B=B_1 e^{ik_FL}-B_2 e^{-ik_FL},\\
    &C=C_1 e^{i\varphi/2}+C_2 e^{-i\varphi/2},\\
    &D_{\pm}=D_1e^{i(\varphi \pm \Delta k_FL)/2}+D_2 e^{-i(\varphi\pm \Delta k_FL)/2}\\
    \end{aligned}
    \label{E4}
\end{equation}
are relevant to the reflection amplitudes in Eq.~(\ref{D18}), with 
\begin{equation}
\begin{aligned}
    A_{\alpha}&=\frac{({k_{F}^{+}}^2-{k_{F}^{-}}^2)\sin^2{\phi_\alpha}}{4k_{F}^{+}k_{F}^{-}+\sin^2{\phi_\alpha}(k_{F}^{+}-k_{F}^{-})^2}
    =\sin^2{\phi_\alpha} \delta, \\
    B_{\alpha}&
    =\frac{2\sqrt{k_{F}^{+}k_{F}^{-}}({k_{F}^{+}}-{k_{F}^{-}})\sin{\phi_\alpha}\cos{\phi_\alpha}}{4k_{F}^{+}k_{F}^{-}+\sin^2{\phi_\alpha}(k_{F}^{+}-k_{F}^{-})^2}
    =\sin{\phi_\alpha}\cos{\phi_\alpha}\delta,\\
    C_{\alpha}&=\frac{4 k_{F}^{+}k_{F}^{-}\cos{\phi_\alpha}}{4k_{F}^{+}k_{F}^{-}+\sin{\phi_\alpha}(k_{F}^{+}-k_{F}^{-})^2}\\
    &=\cos{\phi_\alpha}(1-\sin^2{\phi_\alpha\delta^2}),\\
    D_{\alpha}&=\frac{2\sqrt{k_{F}^{+}k_{F}^{-}}({k_{F}^{+}}+{k_{F}^{-}})\sin{\phi_\alpha}}{4k_{F}^{+}k_{F}^{-}+\sin^2{\phi_\alpha}(k_{F}^{+}-k_{F}^{-})^2}\\
    &=\sin{\phi_\alpha}(1+\frac{\cos{2\phi_\alpha}}{2}\delta^2),\;
    \alpha=1,2.
    \end{aligned}
    \label{E5}
\end{equation}
Here each of the above quantity is kept to the lowest order of $\delta$ in the corresponding second equation, and $\delta=\Delta k_F/(2k_{F})=h/2 E_F$, $\phi_1=\phi_N-\theta_S/2$ (on the left interface) and $\phi_2=\phi_N+\theta_S/2$ (on the right interface). The dimensionless parameter $\delta$ is actually $\sin{ \Theta}$ given by Eq.~(\ref{D15}). Other parameters are also needed to be expanded like $e^{i k_F^{\pm}L}=e^{ik_FL}(1\pm i  \Delta k_F L/2)=e^{ik_FL}(1\pm i k_FL\delta)$. Notice that for the corresponding lattice model, Eq.~(\ref{E3})-(\ref{E5}) are the same except that $\delta$ should be given by Eq.~(\ref{D12}), and $e^{i k_F^{\pm}L}$ should be expanded as $e^{i k_F^{\pm}L}=e^{ik_FL}(1\pm i  \Delta k_F L/2)=e^{ik_FL}(1\pm i L\sin k_F\delta)$ instead.

If $\Delta \varphi$ is of the same order of magnitude as $\delta$, one would arrive at a contradiction. So at least $\Delta \varphi$ is of the order of $\delta^2$, and the expansion of the determinant has to keep the order of $\delta^4$. Now the matrix entries in Eq. (\ref{E4}) can be shown to take the following expansions:
\begin{equation}
    \begin{aligned}
    &A_{\pm}=A_0 \delta \pm iS \chi \delta^2,\;B=B_0\delta,\\
    &C=c_0+\tilde{c}\Delta \varphi+c_2 \delta^2,\;D_{\pm}=d_0+\tilde{d}\Delta \varphi \pm d_1 \delta +d_2 \delta^2,\\
    \end{aligned}
    \label{E6}
\end{equation}
where $\chi=k_F L$ and
\begin{equation}
    \begin{aligned}
    &A_0=-\cos{\chi}\sin2 \phi_N\sin{{\theta_S}}+i \sin{\chi}\left(1-\cos{2 \phi_N\cos{{\theta_S}}}\right),\\
    &S=\cos{\chi}\left(1-\cos{2 \phi_N\cos{{\theta_S}}}\right)-i \sin{\chi}\sin2 \phi_N\sin{{\theta_S}},\\
    &B_0=-\cos{\chi}\cos{2 \phi_N}\sin{{\theta_S}}+i\sin{\chi}\sin{2 \phi_N}\cos{{\theta_S}},\\
    &c_0=-\sin{{\theta_S} e^{-i  \phi_N}},\,\tilde{c}=-\frac{1}{2}(e^{i \phi_N}+\cos{\theta_S} e^{-i \phi_N}),\\
    &c_2=\frac{i}{2}\{-\sin{(2 \phi_N-{\theta_S})}\sin{( \phi_N-{\theta_S}/2)}e^{i{\theta_S}/2}\\
&+\sin{(2 \phi_N+{\theta_S})}\sin{( \phi_N+{\theta_S}/2)}e^{-i{\theta_S}/2}\},\\
    &d_0=ic_0,\,\tilde{d}=\frac{i}{2}\left(e^{i \phi_N}-\cos{{\theta_S}}e^{-i \phi_N}\right),\,d_1=2\chi \tilde{d},\\
    &d_2=-\frac{\chi^2}{2}d_0+\frac{i}{2}\{\cos{(2 \phi_N-{\theta_S})}\sin{( \phi_N-{\theta_S}/2)}e^{i{\theta_S}/2}\\
&-\cos{(2 \phi_N+{\theta_S})}\sin{( \phi_N+{\theta_S}/2)}e^{-i{\theta_S}/2}\}.
\end{aligned}
\label{E7}
\end{equation}
By substituting these expressions into the determinant in Eq.~(\ref{E3}) and keeping to the quartic order of $\delta$ we find the following algebraic equation:
\begin{equation}
    \begin{aligned}
        \delta^4(a_2 {\eta}^2+a_1{\eta}+a_0)=0,
     \end{aligned}
     \label{E8}
\end{equation}
where $\Delta \varphi=\delta^2 \eta$, and the coefficients $a_0$, $a_1$ and $a_2$ for the quadratic polynomial are derived to be given below
\begin{equation}
    \begin{aligned}
    &a_2=4|c_0\widetilde{c}+d_0\widetilde{d}|^2=4\sin^2{{\theta_S}},\\
    &a_1=4 \text{Re}\left\{(c_0 \widetilde{c}+d_0\widetilde{d})^*(2c_0c_2+2d_0d_2-d_1^2)\right\}\\
    &+4\left(|A_0|^2-|B_0|^2\right)\text{Re}\left(d_0\widetilde{d}^*-c_0\widetilde{c}^*\right)\\
    &+8\text{Re}\left(A_0 B_0^*\right)\text{Re}\left(c_0\widetilde{d}^*+d_0 \widetilde{c}^*\right)\\
    &=8\sin{{\theta_S}}(\chi^2-\sin^2{\chi})\left(\cos{2 \phi_N}-\cos{{\theta_S}}\right),\\
    &a_0=\left|2c_0c_2+2d_0d_2-d_1^2\right|^2+\left|A_0^2+B_0^2\right|^2\\
    &+4\left(|A_0|^2-|B_0|^2\right)\text{Re}\left(d_0d_2^*-c_0c_2^*\right)\\
    &-2\left(|A_0|^2+|B_0|^2\right)|d_1|^2
    +8 \text{Re}\left(A_0 B_0^*\right)\text{Re}\left(c_0 d_2^*+d_0c_2^*\right)&\\
    &-2\left(|c_0|^2+|d_0|^2\right)|S|^2\chi^2\\
    &-8\text{Im}\left(d_0d_1^*\right)\text{Re}\left(S A_0^*\right)\chi
    -8\text{Im}\left(c_0d_1^*\right)\text{Re}\left(S B_0^*\right)\chi&\\
    &=4(\chi^2-\sin^2{\chi})^2\left(\cos{2 \phi_N}-\cos{{\theta_S}}\right)^2=\frac{a_1^2}{4a_2}.
    \end{aligned}
    \label{E9}
\end{equation}
Rigorous derivation shows that the polynomial is a perfect square, so the solution of $\eta$ is a repeated root:
\begin{equation}
    \begin{aligned}
\eta=-\frac{a_1}{2a_2}=-\frac{((k_FL)^2-\sin^2{k_FL})\left(\cos{2 \phi_N}-\cos{{\theta_S}}\right)}{\sin{{\theta_S}}}
\end{aligned}.
    \label{E10}
\end{equation}
Physically, the introduction of a Zeeman field can remove the ZEABSs or shift their locations, but cannot create new ZEABSs. So if the solution of $\Delta\varphi$ or $\eta$ exists, due to particle-hole symmetry, it must be unique and so $\eta$ must be a repeated root. If one is aware of this point at the beginning, the above computation can be greatly simplified since the computation of $a_0$ is not necessary. Starting from the other zero-energy level crossing located at $\pi-\theta_s$ we have nearly the same result. Therefore in the presence of a Zeeman field, the new locations of the ZEABSs are $\pm\varphi_0$ with
\begin{equation}
    \begin{aligned}
    &\varphi_0=\pi \\
    &+\left({\theta_S}-\frac{h^2((k_FL)^2-\sin^2{k_FL})\left(\cos{2 \phi_N}-\cos{{\theta_S}}\right)}{4{E_F}^2\sin{{\theta_S}}}\right).
    \end{aligned}
    \label{E11}
\end{equation}
Here $\sin{{\theta_S}}$ must be a finite value. If $\sin{{\theta_S}}$ is so small that it is of the same order of magnitude as $\delta$, the dependence of the the ZEABSs on the Zeeman field will be expected to be very sensitive, and it is hard to obtain an analytical result.   

In the case of the lattice model, as we mentioned before that now $\delta=h/2 \sin^2{k_F}$, we only need to apply the replacement $(k_FL)^2\to(L\sin k_F)^2$ in Eq.~(\ref{E10}). Then the corresponding result for the lattice model can be obtained:
\begin{equation}
    \begin{aligned}
    &\varphi_0=
    \pi\\
    &+\left({\theta_S}-\frac{h^2((L\sin k_F)^2-\sin^2{k_FL})\left(\cos{2 \phi_N}-\cos{{\theta_S}}\right)}{4\sin^4{k_FL}\sin{{\theta_S}}}\right).\\  
    \end{aligned}
    \label{E12}
\end{equation}It is worth noting that as in the case of ${\theta_S}=0$ discussed in Appendix~\ref{D}, the locations of the two level crossings here are still symmetric about $\varphi=\pi$ even if ${\theta_S}\neq 0$. But instead of the linear dependence of $h$ in ${\theta_S}=0$ case, the deviation $\Delta\varphi$ here is proportional to $h^2$, showing a weaker dependence on $h$. In particular, when $\cos{2 \phi_N}=\cos{{\theta_S}}$, namely, when $\phi_N=\pm\theta_S/2$, or $\pi\pm\theta_S/2$, the dependence on $h$ is expected to be even weaker. In this special situation, the Zeeman field is perpendicular to one of the $\boldsymbol{d}$-vectors.
%\end{appendix}
\bibliographystyle{apsrev4-2}
\bibliography{ref}

\begin{thebibliography}{76}
\bibitem [{SM1()}]{SM1}%
  \BibitemOpen
  \href@noop {} {}\bibinfo {note} {For more general
  $\boldsymbol{\mathcal{T}}$-breaking terms such as the hopping intergrals
  being different for two spin species, the two zero-energy level crossings can
  be located non-symmetrically, but they must also be shifted and then removed
  or recreated at the same $\varphi$-position, still exhibiting band-like
  structure.}\BibitemShut {Stop}%
\end{thebibliography}
\end{document}